\documentclass{article}
\usepackage[a4paper, total={6in, 8in}]{geometry}
\usepackage[english]{babel}
\usepackage[utf8]{inputenc}
\usepackage[T1]{fontenc}
\usepackage{setspace}
\usepackage{float}
\usepackage{amsmath}
\usepackage{amssymb}
\usepackage{amsthm}
\usepackage{bbm}
\usepackage{booktabs}
\usepackage{comment}
\usepackage{csquotes}
\usepackage{pdfpages}
\usepackage{lmodern}
\usepackage[most]{tcolorbox}
\usepackage{blindtext}
\usepackage{hyperref}
\usepackage[format=hang, font=it]{caption}
\newtheorem{Definition}{Definition}
\DeclareMathOperator*{\argmin}{arg\,min}
\renewcommand{\epsilon}{\varepsilon}

\usepackage[backend=biber, maxbibnames=10, maxcitenames=2, style=authoryear-comp, natbib]{biblatex}
\title{From Zero-Intelligence to Queue-Reactive: \\ Limit-Order-Book modeling for high-frequency volatility estimation and optimal execution}
\date{\today}
\author{Tommaso Mariotti\\
		Scuola Normale Superiore\\
		\texttt{tommaso.mariotti@sns.it}\\
		\smallskip\\
		Fabrizio Lillo\\
		University of Bologna and Scuola Normale Superiore\\
		\texttt{fabrizio.lillo@unibo.it}\\
		\smallskip\\
		Giacomo Toscano\\
		University of Firenze\\
		\texttt{giacomo.toscano@unifi.it}\\
		\medskip
	}

\begin{document}
\maketitle

\begin{abstract}
{The estimation of the volatility of financial assets with high-frequency data is plagued by the presence of microstructure noise, which  leads to biased measures. Alternative estimators have been developed and tested either on specific structures of the noise or by the speed of convergence to their asymptotic error distributions. \citet{gatheral2010zero} proposed to use the Zero-Intelligence model of the limit order book to test the finite-sample performance of several estimators of the integrated variance. Building on this approach, in this paper we introduce three main innovations: (i) we use as data-generating process the Queue-Reactive model of the limit order book (\citet{huang2015simulating}), which - compared to the Zero-Intelligence model - generates more realistic microstructure dynamics, as shown here by means of the Hausman test by \citet{ait2019hausman}; (ii) we consider not only estimators of the integrated volatility but also of the spot volatility; (iii) we show the relevance of the estimator in the prediction of the variance of the cost of a simulated VWAP execution. In the case of the integrated volatility, we find that the pre-averaging estimator optimizes the estimation bias, while the unified and alternation estimators lead to optimal mean squared error values. Instead, in the case of the spot volatility,  the Fourier estimator  yields  the  optimal accuracy, both in terms of bias and mean squared error. The latter estimator leads also to the optimal  prediction of the cost variance of a VWAP execution.}
\end{abstract}

\section{Introduction}\label{sect:Intro}

The availability of efficient estimates of the volatility of financial assets is crucial for a number of applications, such as  model calibration, risk management, derivatives pricing, high-frequency trading and optimal execution. High-frequency data provide, in principle, the possibility of obtaining very precise estimates of the volatility. The (infill) asymptotic theory of volatility estimators was initially derived under the assumption that the asset price  follows an It\^o semimartingale (see Chapter 3 of \citet{ait2014high}).
The It\^o semimartingale hypothesis ensures the absence of arbitrage opportunities (see \citet{delbaen1994general}) and, at the same time, is rather flexible, as it does not require to specify any parametric form for the dynamics of the asset price.   

However, empirical evidences and theoretical motivations indicate that the prices of financial assets do not conform to the semimartingale hypothesis at high frequencies, due to the presence of microstructure phenomena such as, e.g., bid-ask bounces or price rounding (see  \citet{hasbrouck2007empirical} for a review).   
From the statistical point of view,  such phenomena have been modeled as an additive noise component and  the asymptotic  theory that takes into account the presence of the latter was readily developed (see Chapter 7 of \citet{ait2014high}). 
In its most basic form, the noise due to microstructure is assumed to be i.i.d. and independent of the semimartingale driving the price dynamics (i.e., the so-called efficient price). Moreover,  more sophisticated forms have been studied,  such as,   for instance, an additive noise which is auto-correlated or correlated with the efficient price (see, e.g., \citet{hansen2006realized}). 

The literature on the estimation of the volatility in the presence of noise is very rich. In fact, there exists a number of alternative methodologies making an efficient use of high-frequency prices to reconstruct not only the total volatility accumulated over a fixed time horizon, i.e., the integrated volatility, but also the trajectory of the latter on a discrete grid, i.e., the spot volatility. 
These include the two-scale and multi-scale approach by, respectively,  \citet{zhang2005tale} and \citet{zhang2006efficient}, the kernel-based method, originally proposed in \citet{barndorff2008designing},    the Fourier-transform method  by \citet{malliavin2002fourier,malliavin2009fourier},  and the pre-averaging approach by \citet{jacod2009microstructure}. Given this variety of alternative methodologies, it is not straightforward to establish which specific noise-robust estimator should  be preferred for high-frequency financial applications.

As pointed out in the seminal paper by \citet{gatheral2010zero}, the best asymptotic properties, i.e., the optimal rate of convergence and the minimum asymptotic error variance, do not guarantee the best performance in finite-sample applications.  \citet{gatheral2010zero} proposed  to compare the finite-sample performance of different high-frequency estimators via simulations based on a market simulator which is able to reproduce the actual mechanism of price formation at high frequencies  with sufficient realism. In this regard, the authors used simulations obtained via the Zero-Intelligence (ZI) limit order book  model by \citet{smith2003statistical} to compare the performance of different integrated volatility estimators. However, the ZI model is based on several simplistic assumptions on the dynamics of the limit order book, and thus it may fail to replicate the actual behavior of high frequency financial data and microstructure noise with satisfactory accuracy.  For example, under the ZI model, the order flow  is described by independent Poisson processes, while it is well-known that the order flow is a long-memory process (see \citet{lilloFarmer2004}) and the different components of the order flow are lead-lag cross-correlated (\citet{eisler}). Moreover, as pointed out by \citet{bouchaud2018trades}, the ZI model leads to systematically profitable market making strategies. These properties are likely to have an effect on the dynamics of the volatility and  market microstructure noise, and thus an analysis based on a more realistic limit order book model is needed.  

The first goal of this paper is to extend the study by \citet{gatheral2010zero} in two directions. First, we use a  more realistic limit order book model, namely the Queue-Reactive (QR) model by \citet{huang2015simulating}. Under this model, the arrival rates of orders depend on the state of the limit order book. This implicitly introduces  auto- and cross-correlations of the book components, thereby generating more realistic   dynamics for the price process at high-frequencies. Secondly, we compare not only the performance of a number of estimators of the integrated volatility  (expanding the collection of estimators considered in the study by \citet{gatheral2010zero}), but    also   that of different estimators of the spot volatility. To make our comparison meaningful for applications, the performance of the estimators is evaluated in terms of the  optimization of the bias and the mean-squared-error via the feasible selection of the tuning parameters involved in their implementation.
 Note that, following \citet{gatheral2010zero}, we consider three alternative price series for the estimation: the mid-price, that is, the average between the best bid and best ask quotes; the micro-price, i.e., the volume-weighted average of the best bid and best ask quotes; the trade price, namely the price at which a market order is executed.  
 
For what concerns the integrated variance, we find that the pre-averaging estimator by \citet{jacod2009microstructure} is favorable in terms of bias minimization. Instead, when looking at the optimization of the mean squared error, the situation appears to be more nuanced. Indeed, the   Fourier estimator by \citet{malliavin2009fourier} obtains the best average ranking across the considered price series  (mid-, micro- and trade- prices) without actually achieving the best ranking for any of these individual series. The best rankings are instead achieved by the unified volatility estimator by \citet{li2018unified} (for mid- and micro-prices) and by the alternation estimator by \citet{large2011estimating} (for trade-prices).
Instead, for what concerns the spot variance,  the Fourier estimator provides the relative best performance for the three prices series, both in terms of bias and mean-squared-error optimization.

The second goal of the paper is to study the impact of the availability of efficient volatility estimates on  optimal execution.  Specifically, we investigate, via simulations of the QR model, how the use of different volatility estimators affects the inference of the variance of the cost of the execution strategy. To do so, we consider the instance where the trader is set to execute a volume-weighted average price (VWAP) strategy and assumes that market impact is described by the Almgren and Chriss (\citet{almgren2001optimal}) model. 
We compare the empirical variance of the implementation shortfall of the simulated executions with the corresponding model-based prediction, evaluated with different spot volatility estimators. As a result, we find that the estimator that yields the optimal performance in terms of bias and mean-squared-error optimization, namely the Fourier estimator,  also gives the optimal forecast of the cost variance. More generally, our results suggest that the choice of the spot estimator is not irrelevant, as it may lead to significantly different forecasts of the variance of the implementation shortfall.

The paper is organized as follows. In Section \ref{sect:LOBmodels}  we recall the main characteristics of the ZI and QR limit-order-book models, discuss their calibration on empirical data and   compare their ability to reproduce realistic  volatility and noise features. In Section \ref{sect:estimators}  we illustrate the estimators of the integrated and spot variance, while in Section \ref{sect:performanceCOMP} we evaluate their finite-sample performance  with simulated   data from the QR model.  {Finally,   Section \ref{sec:optex} contains the study of the impact of efficient volatility estimates on optimal execution.} Section \ref{sect:conclusions} concludes.

\section{Limit-order-book models: zero-intelligence vs queue-reactive}\label{sect:LOBmodels}

Electronic financial markets are often based on a double auction mechanism, with a bid (buy) side  and an  ask (sell) side. The limit order book (LOB) is the collection of all the outstanding limit orders, which are orders of buying or selling a given quantity of the asset at a given price, expressed as a multiple of the tick size (i.e., the minimum  price movement allowed) of the asset. Other two types of orders can be placed: a cancellation, that erases a limit order previously inserted by the same agent, thereby reducing the volume at a given price level, and a market order, that is, an order to immediately buy/sell the asset at the best possible price.
The best bid is the highest price at which there is a limit order to buy, and the best ask is the lowest price at which there is a limit order to sell. The spread is the difference between the best ask and the best bid, and is typically expressed in tick size. For a detailed overview of the LOB see \citet{abergel2016limit}. 

In the following, we will be interested in   three  price series that can be retrieved from LOB data: the mid-price, the micro-price and the trade price. 

\begin{Definition}
We define the mid-price $p_{mid}$ and the micro-price $p_{micro}$ of an asset at time $t$ as, respectively, the arithmetic average and the volume-weighted average of the best bid and best ask quotes at time $t$, i.e.,

$$p_{mid}(t) := \frac{p^b(t)+p^a(t)}{2}, \qquad  p_{micro}(t):= \frac{p^b(t)v^a(t)+p^a(t)v^b(t)}{v^b(t)+v^a(t)},$$

\noindent where $p^b$, $p^a$, $v^b$ and $v^a$ denote, respectively, the best bid, the best ask, the volume (i.e., the number of outstanding limit orders) at the best bid and the volume at the best ask. Finally,
 the trade price $p_{trade}$ series is defined as the series of prices arising from the execution of market orders.
\end{Definition}

In our study, we will consider two models for the simulation of the LOB. The simplistic ZI model by \citet{smith2003statistical} and the more sophisticated QR model by \citet{huang2015simulating}. In the next subsections we briefly recall the main characteristics of the two models. Please refer to the original papers for a more thorough description.

\subsection{Model descriptions}

\subsubsection*{The zero-intelligence model}
The ZI model, originally proposed by \citet{smith2003statistical}, is a statistical representation of the double action mechanism used in most stock markets. Despite its simplicity, the model is able to generate a relatively complex dynamic for the order book. It is based on three parameters: the intensity of limit orders, $\lambda^L$, the intensity of cancel orders, $\lambda^C$,  and the intensity of market orders, $\lambda^M$. The three components of the order flow follow independent Poisson processes, thus the type of order extracted at each time is independent of the previous orders and the current state of the LOB, and orders may arrive at every price level with the same probability. Each order is assumed to have unitary size. For a detailed discussion about the flexibility of this model, see \citet{gatheral2010zero}.

As mentioned, the ZI model may be deemed as too simplistic. Indeed, the assumptions that the intensities of order arrival are independent of the state of the book and that the intensities are equal for each price level are highly unrealistic. Moreover, this model produces purely endogenous order-book dynamics, without considering the effect of exogenous information. Further, as shown in  \citet{bouchaud2018trades}, under the ZI model the market impact of new orders is such that profitable market-making opportunities can be created, even if they are usually absent in real markets. Some of the weakness of the ZI model are overcome by the QR model.

\subsubsection*{The queue-reactive model}
The QR model (\citet{huang2015simulating}) is a LOB model suitable to describe large tick assets, i.e., assets whose bid-ask spread is almost always equal to one tick. 
This model is able to reproduce a richer and more realistic behavior of the LOB, compared to the ZI model. In other words, the QR model attempts to fix some of the flaws of the ZI model. This is achieved, in the first place,  by assuming different intensities for each level of the LOB. Moreover, the degree of realism is increased  by introducing a correlation not only between order-arrival intensities and the corresponding queue size at each level, but also between    intensities and the queue size at the corresponding level at the opposite side of the book. Further, a dependence between the volume at the best level and  order arrivals at the other levels is assumed. 
Finally, differently form the ZI model, the QR model allows for exogenous dynamics by taking into account the flow of exogenous information that hits the market.

Following \citet{huang2015simulating}, under the QR model the LOB is described by a $2K-$dimensional vector, with $K$ denoting the number of available price levels at the bid and ask sides of the book. At the level  $Q_{\pm i}, \, i = 1,...,K$, the corresponding price is equal to $p\pm i(tick)$, where $p$ denotes the center of the $2K$ dimensional vector. Precisely, $Q_{-i}$ denotes a level order at the bid side and $Q_i$ denotes a level order at the ask side. Moreover, $q_{\pm i}$ denotes the volume at the level $Q_{\pm i}$. 

The process $X(t) = (q_{-K}(t),...,q_{-1}(t),q_1(t),..., q_K(t))$ is a continuous-time Markov process with the following infinitesimal generator matrix $\mathcal{Q}$:
\begin{align*}
    & \mathcal{Q}_{q,q+e_i} = f_i(q),\\
    & \mathcal{Q}_{q,q-e_i} = g_i(q),\\
    & \mathcal{Q}_{q,q} = -\sum_{p \neq q} \mathcal{Q}_{q,p},\\
    & \mathcal{Q}_{q,q} = 0 \quad \text{otherwise},
\end{align*}
with $q=(q_{-K},...,q_{-1},q_1,...,q_K)$  and $e_i$ denoting the $i-th$ vector of the standard base of $\mathbb{R}^{2K}$. 

Thus, with different intensities at each queue, we have that
\begin{align*}
    & f_i(q) = \lambda^L(q_i, S_{m,l}(q_{-i})),\\
    & g_i(q) = \lambda^C(q_i, S_{m,l}(q_{-i})) + \lambda^M_{\text{buy}}\mathbbm{1}_{\text{bestask}(q) = i}, \quad \text{if} \quad i>0,\\
    & g_i(q) = \lambda^C(q_i, S_{m,l}(q_{-i})) + \lambda^M_{\text{sell}}\mathbbm{1}_{\text{bestbid}(q) = i}, \quad \text{if} \quad i<0,
\end{align*}
where the function $S_{m,l}(q_{-i})$ is responsible for the interaction between the bid and the ask side of the book, that is,
\begin{align*}
    S_{m,l}(q_{-i}) &= Q^0 \quad \text{if} \quad q_{-i} = 0, \\
    S_{m,l}(q_{-i}) &= Q^- \quad \text{if} \quad 0 < q_{-i} \leq m, \\
    S_{m,l}(q_{-i}) &= \bar{Q} \quad \text{if} \quad m < q_{-i} \leq l,\\
    S_{m,l}(q_{-i}) &= Q^+ \quad \text{if} \quad q_{-i} > l,
\end{align*}
being $m$ and $l$ two fixed thresholds. For example, given a certain volume at the bid side, a new bid limit order has different intensities depending on whether the volume at the ask is, e.g., $q_{i} = 0$, $0 < q_i \leq 5$, $5 < q_i \leq 10$ or $q_i > 10$. Market orders may arrive only at the best quote.   Moreover, we assume that $\lambda^L_i$ and $\lambda^C_i$ are also functions of $\mathbbm{1}_{q_{\pm 1} \ge 0}$, for $i \neq \pm 1$, to allow for interactions between the best level and the   dynamics far from the best level.

Conditionally on the LOB state, the arrival of different orders at a given limit is assumed to be independent, and follows a Poisson distribution, with intensity equal to $\lambda$. However, since the queue sizes depend  on the order flow, the model reproduces some auto- and cross-correlations between the components of the order flow, as observed in empirical data. 

Contrary to the ZI model, the QR gives a dynamic that is not entirely driven by the orders arriving on the LOB. To achieve this, two additional parameters are introduced, $\theta$ and $\theta^{\text{reinit}}$. 
Whenever the mid-price changes,    an auxiliary (not observed) price,  called the reference price and denoted by $p_{ref}$, changes in the same direction  with  probability $\theta$ and by an amount equal to the tick size of the asset. Moreover,  when $p_{ref}$ changes, the LOB state is redrawn from its invariant distribution   around the new value of $p_{ref}$,     with   probability  $\theta^{\text{reinit}}$. In fact, as the authors explain, the parameter  $\theta^{\text{reinit}}$ captures
 the percentage of price changes due to exogenous information.

In the next subsection we address the  procedures that we followed to estimate the two LOB models and discuss the differences between them in terms of ability of reproducing realistic features of empirical high-frequency data.

\subsection{Calibration procedures}

For the calibration of the ZI and QR models, we used order-book data of the stock Microsoft (MSFT) over the period April 1, 2018 - April 30, 2018. 
Data were retrieved from the LOBSTER database. Microsoft is a very liquid stock with an average spread approximately equal to $1.25$ ticks and thus can be considered a large tick asset, suitable to be modeled by the queue-reactive design.

Before the calibration, for each day we removed from the sample the first hour of trading activity after the market opening and the last 30 minutes before the market closure; this is a standard procedure adopted when working with high-frequency data, since during these two moments of the day the trading activity is known to be more intense and volatile, thereby possibly leading to a violation of the large tick asset hypothesis, even for a liquid stock like Microsoft. Given the average spread observed, and being the activity almost fully concentrated at the best limits, we  implemented the ZI and  QR models using two limits, $Q_{\pm 1}$ and $Q_{\pm 2}$. This is in line with \citet{huang2015simulating}.  

To estimate the intensities of order arrivals under the QR model, the following inputs are needed:
\begin{itemize}
    \item the type of each event, i.e., limit order, cancel order or market order;
   \item the time between events that happen at $Q_1$ and $Q_2$, along with the queue sizes $q_1$ and $q_2$ before each event;
    \item the size of each event, as $q_i$ is expressed as a multiple of the median event size. 
\end{itemize}

The estimation of the intensities is  performed via maximum likelihood, as in \citet{huang2015simulating}. The parameters $m$ and $l$ that capture the bid-ask dependence are set equal to the the 33\% lower and upper quantiles of the $q_{-i}$'s (conditional on positive values). Given the symmetry property of the LOB, intensities are computed for just one side.
 
 The parameters $\theta$ and $\theta^{\text{reinit}}$ are calibrated using the mean-reversion ratio $\zeta$ of the mid-price, which is  defined as 
 \[
 \zeta := \frac{n_c}{2n_a},
 \]
 where $n_c$ is the number of continuations (i.e., the number of consecutive price moves in the same directions) and $n_a$ is the number of alternations (i.e., the number of consecutive price moves in opposite directions). For more details about the relation between the mean-reversion ratio and the microstructure of large tick assets, see \citet{robert2011new}.  
 
 We carried out the calibration using a two-step generalized method of moments (GMM), which is more robust than the heuristic approach proposed by \citet{huang2015simulating}. Denote by $\sigma^{emp}$ and $\zeta^{emp}$ the empirical estimates of the standard deviation and mean-reversion ratio of the mid-price returns, computed at the 1-second frequency using the last tick rule. Further, denote by  $\sigma_t(\bar{\theta})$ and $\zeta_t(\bar{\theta})$ the quantities estimated in simulation $t$, with $t = 1, ..., T$, and $\bar{\theta} = (\theta,\theta^{\text{reinit}})$.
 
 \vspace{0.5cm}
 
\begin{tcolorbox}[colframe=white]
\textbf{Two-step GMM-based procedure for the calibration of} $\bar\theta$.\\

\textbf{Step 1.} Obtain a consistent estimate of $\bar\theta$ via the estimator

 \[
 \bar{\theta}_1 := \argmin_{\bar{\theta}\in [0,1]\times [0,1]} \frac{1}{T^2} \left[ \left(\sum_t g_t^{\sigma}(\bar{\theta})\right)^2 + \left(\sum_t g_t^{\zeta}(\bar{\theta})\right)^2\right],
 \] 
 where  $\displaystyle g_t^{\sigma}(\bar{\theta}) = \frac{\sigma_t(\bar{\theta})}{\sigma^{emp}} - 1, \, \, \,  g_t^{\zeta}(\bar{\theta}) = \frac{\zeta_t(\bar{\theta})}{\zeta^{emp}} - 1.$\\

\textbf{Step 2.} Obtain the GMM-estimate of $\bar \theta$ 
 \[
 \theta_{GMM} := \argmin_{\bar{\theta}\in [0,1]\times [0,1]} \left( \frac{1}{T}\sum_t  \bar{g}_t(\bar{\theta}) \right)' W \left( \frac{1}{T}\sum_t  \bar{g}_t(\bar{\theta}) \right),
 \]
 where $\displaystyle \bar{g}_t(\bar{\theta}) = (g_t^{\sigma}(\bar{\theta}), g_t^{\zeta}(\bar{\theta}))$ and  $\displaystyle W^{-1} = \frac{1}{T} \sum_t (\bar{g}_t(\bar{\theta}_1))'(\bar{g}_t(\bar{\theta}_1)).$
\\
\end{tcolorbox}

The estimator is asymptotically efficient in the GMM class.  For the implementation we used $T=100$ simulations with an horizon of one trading day. 
The tick size was set equal to the minimum bid-ask spread recorded in the data, namely 1 cent.
As a final estimate, we obtained $\theta=0.6$ and $\theta^{\text{reinit}}=0.85$.

The asymptotic distribution of the queue size, needed for the re-initialization of the LOB state, was obtained following  the approach proposed by \citet{huang2015simulating}. For each simulated path, the starting LOB state was randomly chosen using the asymptotic distribution of the queue size.

For the calibration of the ZI model, one only 
needs to reconstruct the intensities of order arrivals. To do that, the only information needed involves the type of order, the order arrival time and the order size. The ensuing estimators read
\[
\lambda^L = \frac{\#L}{\#O\overline{\Delta t}}, \quad \lambda^C = \frac{\#C}{\#O\overline{\Delta t}}, \quad \lambda^M = \frac{\#M}{\#O\overline{\Delta t}}
\]
where $\#O= \#L+\#C+\#M$ and $\#L$, $\#C$, $\#M$ denote, respectively, the total number of limit, cancel and market orders arriving at the best quotes or between the spread, while $\overline{\Delta t}$ is the average elapsed time between two consecutive orders.
 
\subsection{Comparison of volatility and noise features}\label{features}
 
As pointed out by \citet{gatheral2010zero}, neither the efficient price nor its volatility are well-defined under the ZI model. The same holds under the QR model. However,  if one assumes constant model parameters, thanks to the ergodicity of the processes (see \citet{huang2015simulating}), the variance of the efficient price can be defined, following \citet{gatheral2010zero}, as 

\begin{equation}\label{truevol}
\sigma^2:=\lim_{m\to \infty} \frac{1}{m}E\left[\left(p(m)-p(0)\right)^2 \right],     
\end{equation}
where $p$ may denote any price process among those considered, that is, the mid-price, the micro-price and the trade price.
 
Based on (\ref{truevol}), given the (calibrated) values of the LOB parameters, it is possible to estimate the true value of $\sigma^2$ via simulations. Specifically, in our study, given the LOB parameters calibrated on the Microsoft sample data, numerical results over 2500 simulations show that for $m>18000$ the volatility of mid-prices, micro-prices and trade-prices stabilizes around the value $\sigma^2=1.039 \cdot 10^{-8}$ in the case of the QR model.

Since the knowledge of the value of $\sigma^2$ is crucial to compare the finite-sample performance of volatility estimators, we wish to verify whether the two order-book models give similar results. The estimates of $\sigma^2$ for the two models and for the three price series considered are compared in Table \ref{tab:volconv}. The levels of variance obtained with the ZI model and the QR are significantly different, even when considering the error in the estimation procedure, with the ZI producing a level about 50\% higher than the one observed for the QR model. This highlights a first fact to be considered: the choice of the LOB model is not irrelevant for applications involving the volatility parameter. In fact, it is known that the ZI model calibrated on real data only partially reproduces the actual empirical variance, with a bias which depends on the relative magnitude of the intensities (see, e.g., \cite{bouchaud2018trades}).

\begin{table}[htbp!]
	\begin{center}
		\begin{tabular}{l|c|c|c|c}
	 	    \hline\hline
			Order-book model& statistics & mid-price & micro-price & trade price \\
			\hline\hline
			ZI model &  Variance   & $1.539\cdot 10^{-8}$ & $1.539\cdot 10^{-8}$ & $1.539\cdot 10^{-8}$\\
			& Std. error	& $1.259\cdot 10^{-9}$ & $2.434\cdot 10^{-9}$ & $2.587\cdot 10^{-9}$\\
			\hline
			 QR model &  Variance   & $1.039 \cdot10^{-8}$ & $1.039 \cdot 10^{-8}$ & $1.039\cdot 10^{-8}$\\
			& Std. error	& $3.336\cdot 10^{-10}$ & $3.330\cdot 10^{-10}$ & $3.319\cdot 10^{-10}$\\
		\end{tabular}
		\caption{Comparison of the estimated variance reconstructed via Eq. (\ref{truevol}) using   simulations of the ZI and the QR models.}
		\label{tab:volconv}
	\end{center}
\end{table}

Moreover, since we are interested in comparing the performance of noise-robust volatility estimators, we wish to discriminate between the two LOB models on the basis of how well they mimic the noise accumulation observable in empirical data at different frequencies. To do so, we use the Hausman test for the null hypothesis of the absence of noise by \citet{ait2019hausman}. In particular, we use the formulation of the test in Equation (16) of \citet{ait2019hausman}, which is coherent with the use of  LOB models  with a constant variance parameter.  

Tables \ref{tab:noisetestMSFT}, \ref{tab:noisetestZI} and \ref{tab:noisetestQR} illustrate the frequencies (in seconds) at which the Hausman test rejects the null hypothesis of the absence of noise with a significance level of 5\% ($\bigstar$)  and the frequencies at which the null is instead not rejected (\textdied)  for, respectively, the MSFT sample and the simulated samples from the ZI and QR models. The  results of Hausman test suggest that the noise accumulation mechanism at different frequencies under the QR model is more realistic than the one observed under the ZI model, based on the comparison with the noise-detection pattern in the MSFT sample. This aspect is   clearly relevant when analyzing the finite-sample performance of noise-robust volatility estimators, and adds empirical support to the use of the QR model for that purpose.

\begin{table}[htbp!]
	\begin{center}
		\begin{tabular}{l|c|c|c|c|c|c|c}
		    \hline\hline
			& 1 sec. & 2 sec. & 5 sec. & 10 sec. & 15 sec.& 30 sec.& 60 sec.\\
			\hline\hline
			Mid-price & $\bigstar$ &  $\bigstar$ &  $\bigstar$ & $\bigstar$ &  $\bigstar$ & \textdied & \textdied \\
			Micro-price & $\bigstar$ &  $\bigstar$ &  $\bigstar$ & $\bigstar$ &  $\bigstar$ & \textdied & \textdied \\
			Trade-price &  $\bigstar$ &  $\bigstar$ &  $\bigstar$ & $\bigstar$ &  $\bigstar$ & $\bigstar$ & \textdied
		\end{tabular}
		\caption{Hausman  test results at different frequencies (in seconds) for Microsoft (April 2018). The symbol $\bigstar$ (\textdied) indicates that the null of absence of noise is rejected (not rejected) with a significance level of 5\%. }\label{tab:noisetestMSFT}
	\end{center}
\end{table}

\begin{table}[htbp!]
	\begin{center}
		\begin{tabular}{l|c|c|c|c|c|c|c}
		    \hline\hline
			& 1 sec. & 2 sec. & 5 sec. & 10 sec. & 15 sec.& 30 sec.& 60 sec.\\
			\hline\hline
			Mid-price & $\bigstar$ & $\bigstar$ & $\bigstar$ & $\bigstar$ & \textdied & \textdied &\textdied\\
			Micro-price & $\bigstar$ & $\bigstar$ & $\bigstar$ & \textdied & \textdied & \textdied &\textdied \\
			Trade-price & $\bigstar$ & $\bigstar$ & $\bigstar$ & \textdied & \textdied & \textdied &\textdied
		\end{tabular}
		\caption{Hausman  test results at different frequencies (in seconds) for the ZI model. The symbol $\bigstar$ (\textdied) indicates that the null of absence of noise is rejected (not rejected) with a significance level of 5\%.}\label{tab:noisetestZI}
	\end{center}
\end{table}

\begin{table}[htbp!]
	\begin{center}
		\begin{tabular}{l|c|c|c|c|c|c|c}
		    \hline\hline
			& 1 sec. & 2 sec. & 5 sec. & 10 sec. & 15 sec.& 30 sec.& 60 sec.\\
			\hline\hline
			Mid-price & $\bigstar$ &  $\bigstar$ &  $\bigstar$ & $\bigstar$ &  $\bigstar$ & \textdied & \textdied \\
			Micro-price & $\bigstar$ &  $\bigstar$ &  $\bigstar$ & $\bigstar$ &  $\bigstar$ & \textdied & \textdied \\
			Trade-price & $\bigstar$ &  $\bigstar$ &  $\bigstar$ & $\bigstar$ &  $\bigstar$ & \textdied & \textdied
		\end{tabular}
		\caption{Hausman  test results at different frequencies (in seconds) for the QR model. The symbol $\bigstar$ (\textdied) indicates that the null of absence of noise is rejected (not rejected) with a significance level of 5\%.}\label{tab:noisetestQR}
	\end{center}
\end{table}

Lastly, we look at the average spread that the two order-book models are able to generate. This aspect is of paramount importance, being the spread a crucial characteristic of LOB   and one of the main sources of market microstructure noise.  Table \ref{tab:spread} suggests that, even though both    the ZI and the QR models generate an average spread   lower than the one empirically observed, the underestimation is less severe in the case of the QR. The underestimation of spread under the ZI model has been documented in \citet{bouchaud2018trades}.  

\begin{table}[htbp!]
	\begin{center}
		\begin{tabular}{c|c}
	    	\hline\hline
		    Data & Average spread\\
		    \hline\hline
		    MSFT sample & 1.25\\
			ZI simulations & 1.03\\
			QR simulations & 1.14 \\
		\end{tabular}
		\caption{Comparison of the average spread (expressed in ticks) computed from the empirical MSFT sample with the corresponding values obtained from simulations of the ZI and QR samples.}\label{tab:spread}
	\end{center}
\end{table}

\section{Volatility estimators}\label{sect:estimators}
In this section, we briefly describe the noise-robust integrated- and spot-volatility estimators whose performance will be studied in the next section.  The formulae for the tuning parameters involved in the computation of the estimators that optimize the mean squared error (MSE) are  also reported.
 
\subsection{Preliminary notation}

We consider the estimation horizon $[t,t+h]$, $t,h>0$,  and assume that the price   $p$ is sampled on the equally-spaced grid with mesh $h/n$, where $n$ denotes the number of price  observations. The quantity $p_i$ denotes the log-price of the asset at time $t_i:=t+ih/n$, $i=0,1,...,n$. Further, we define $\Delta p_i := p_i -p_{i-1}$. Note that $p$ may refer indifferently to the trade-, mid- or  micro-price.

The spot volatility at time $t$ is denoted by $\sigma^2(t)$ and 

$$IV_{t,u}:=\int_t^{t+u} \sigma^2(s) ds$$ denotes the integrated volatility on  $[t,t+u]$, $u\le h$. Clearly, in a setting with constant volatility $\sigma^2(t)=\sigma^2$ for all $t$, the latter simplifies to $\sigma^2u$. 

As  an auxiliary quantity for the implementation of some estimators,  we will  need to estimate the integrated quarticity, that is,  

$$IQ_{t,u}:=\int_t^{t+u} \sigma^4(s) ds.$$  

In the rest of the paper, we drop the subscript of both $IV$ and $IQ$ as we always refer to the interval $[t,t+h]$.
 
Furthermore, we recall that the asymptotic properties of high-frequency volatility estimators are typically derived under the assumption that, for all $t$, the observable price $p$ is decomposed as 

$$p(t) = p^{eff}(t) + \eta(t), $$
where  $p^{eff}$ denotes the efficient price, whose dynamics follow an It\^o semimartingale, while  $\eta$ is an i.i.d. zero-mean noise due to the market microstructure. As additional auxiliary quantities, we will need estimates of the second moment of $\eta$, i.e.,
  
\[\omega^2:=E[\eta^2]\]

Finally, we denote the floor function as $\lfloor \cdot \rfloor$ and the rounding to the nearest integer as $\left[ \cdot\right]$.

\subsection{Integrated volatility}
 
\subsubsection*{Bias-corrected realized variance}
The realized variance, that is, the sum of squared log-returns over a given time horizon, represents the most natural  rate-efficient estimator of the integrated volatility in the absence of noise. However, in the presence of noise, as it is typically the case for high-frequency settings, the realized variance is biased.   The bias-corrected realized variance by \citet{zhou1996high} corrects for the bias due to noise  by taking into account the first order auto-covariance of the log-returns. The estimator reads:
\[
IV_{BC} := c\sum_{j=0}^{q-1}\left(\sum_{i=1}^{c_2}(p_{iq+j}-p_{(i-1)q+j}\right)^2+2\sum_{i=1}^{c_2-1}(p_{iq+j}-p_{(i-1)q+j}) (p_{(i+1)q+j}-p_{iq+j})  ,
\]
where $c = \frac{n}{n-q+1}\frac{1}{q}$ and $c_2= \lfloor (n-j+1)/q\rfloor$.
    
The MSE-optimal value of q is attained as 
\[
q^* = \max \left(1, \left[\frac{2n \omega^2}{IV\sqrt{3}}\right]\right).
\]

\subsubsection*{Fourier estimator}
 
 Introduced by \citet{malliavin2002fourier,malliavin2009fourier}, the Fourier estimator of the integrated volatility relies on the computation of the zero-th Fourier coefficient of the volatility, given the Fourier coefficients of the log-returns. The noise is filtered out by   suitably selecting   the cutting frequency $N$. If one uses the Fej\'er (respectively, Dirichlet) kernel to weight the convolution product, the estimator is defined as   

\[
IV^{Fej}_F:= \frac{(2\pi)^2}{ N+1}\sum_{|k| \le N} \left(1-\frac{|k|}{N+1}\right) c_k(dp_n)c_{-k}(dp_n),  
\]
and 
\[
IV^{Dir}_F:= \frac{(2\pi)^2 }{ 2N+1}\sum_{|k| \le N}   c_k(dp_n)c_{-k}(dp_n),  
\]
where 

$$c_k(dp_n)=\frac{1}{2\pi}\sum_{j=1}^{n}\textrm{e}^{-\textrm{i}kt_j}\Delta p_j$$ 
represents the k-th discrete Fourier coefficient of the log-return.

The optimal value of the integer $N$ in the presence of noise can be selected by performing a feasible minimization of the MSE, see \citet{mancino2008robustness}.

In this paper, we implemented  $IV^{Fej}_F$, as unreported simulations suggest that it  performs  better than  $IV^{Dir}_F$.  
 
\subsubsection*{Maximum likelihood estimator}

 The maximum-likelihood estimator by \citet{ait2005often} is based on the assumption that noisy log-returns follow  an MA(1) model, consistently with the seminal microstructure model  for the bid-ask spread by \citet{roll1984simple}. under the MA(1) assumption, it holds that 
\begin{equation}\label{MLEnoise}
\Delta p_i = w_i +\phi w_{i-1},    
\end{equation}
 
and the maximum-likelihood  estimator reads
\[
IV_{ML} := n \hat{\sigma}^2_{w} (1+\hat{\phi})^2,
\]
where the pair $(\hat{\phi}, \hat{\sigma}^2_{w})$ is the result of the standard maximum-likelihood estimation of the MA(1) model.

\subsubsection*{Two-scale estimator}
The two-scale realized variance by \citet{zhang2005tale} eliminates the noise-induced bias of the realized variance by combining two different realized variance values, one computed at a higher frequency and one computed at a lower frequency. The estimator reads:

\[
IV_{TS} := c\left(\frac{1}{q}\sum_{j=0}^{q-1}\sum_{i=1}^{\lfloor (n-j+1)/q\rfloor}(p_{iq+j}-p_{(i-1)q+j})^2 -\frac{n-q+1}{nq} \sum_{i=1}^{n}\Delta p_i^2\right),
\]
where $c=\left(1-\frac{n-q+1}{nq}\right)^{-1}$. The MSE-{optimal} value of $q$ is equal to 
 
\[
q^*= n^{2/3}\left(\frac{12\omega^4 }{IQ}\right)^{1/3} .
\]

\subsubsection*{Multi-scale estimator}
\citet{zhang2006efficient} also proposed a more sophisticated combination of realized variances at various frequencies that smooths out the effect of microstructure noise. The multi-scale estimator reads:
    
\[
IV_{MS} := \sum_{j=1}^q \frac{a_j}{j} \sum_{k=0}^{j-1}\sum_{i=1}^{\lfloor (n-k+1)/j\rfloor}(p_{ij+k}-p_{(i-1)j+k})^2,
\]
where
\[
a_j = j(1-1/q^2)^{-1}\left(\frac{12(1/q -1/2)}{q^2}-\frac{6}{q^3}\right).
\]
The MSE-{optimal} $q$ is given by 
\[ q^* = \sqrt{n}\left(\beta +\sqrt{\beta^2+\frac{144\omega^4}{104/35IQ}}\right)^{1/2},
\]
where $\beta = \displaystyle \frac{48/5\omega^2(IV+\omega^2/2)}{208/35IQ}$.

\subsubsection*{Kernel estimator}
Kernel-based estimators, originally introduced by \citet{barndorff2008designing}, correct for the bias due to noise of the realized variance by taking into account the autocorrelation of returns at different lags, suitably weighted by means of a kernel function $k(\cdot)$.  The estimator reads:

\[IV_{K} := \sum_{i=1}^{n}\Delta p_i^2+2\sum_{j=1}^q k\left(\frac{j-1}{q}\right)\sum_{i=1}^{n-j}\Delta p_i \Delta p_{i+j}.
\]
Moreover, the MSE-{optimal} value of $q$ is equal to
\[
q^* =  \sqrt{n}\left(\beta +\sqrt{\beta^2+\frac{c\omega^4}{aIQ}}\right)^{1/2} ,
\]
where $\beta = \displaystyle \frac{b\omega^2(IV+\omega^2/2)}{2aIQ}, \, a = 4\int_0^1 k(x)^2dx, \, b= -8\int_0^1k(x)k^{''}(x)dx, \, c=12(k^{'''}(0)+\int_0^1k(x)k^{'''}(x)dx$.

In this paper we implemented this estimator by using the Tukey Hanning 2 kernel, i.e., we set $k(x) = \sin^2(\pi/2(1-x)^2)$. This kernel was shown to perform satisfactorily, compared to other kernels (see \citet{barndorff2008designing}).
 
\subsubsection*{Pre-averaging estimator}
 The pre-averaging estimator, proposed by \citet{jacod2009microstructure}, relies on the averaging of the price values over a window $h$ to compute the realized variance, together with a bias-correction term. The estimator is as follows:
    
\[
IV_{PA} := \frac{6}{h}\sum_{s=0}^{n-h+1}\left(\frac{1}{h/2}\sum_{s'=0}^{h/2-1}p_{s+s'+h/2}
-\frac{1}{h/2}\sum_{s'=0}^{h/2-1}p_{s+s'}\right)^2-\frac{6}{h^2}\sum_{i=1}^{n}\Delta p_{i}^2.
\]

\citet{jacod2009microstructure} suggest that the estimator is robust to the choice of $h$;  in our simulation study we set $h$ to obtain a window of approximately 4 minutes, in line with \citet{li2021volatility}.

\subsubsection*{Alternation estimator}
Proposed by \citet{large2011estimating}, the alternation estimator corrects the  realized variance with a factor dependent on the number of alternations and continuations in the sample, that is, the estimator reads
  
\[
IV_{Alt}:=\frac{n_c}{n_a}\sum_{i=1}^{n}\Delta p_i^2,
\]
where $n_c$ is the number of consecutive price movements in the same direction, while $n_a$ the number of consecutive price movement in the opposite direction.

\subsubsection*{MinRV and MedRV estimators}
\citet{andersen2012jump} introduced two jump-robust estimators of the integrated variance  which consist, respectively, in the (scaled) sum of the minimum or the median between consecutive returns, that is,

\[
IV_{Min}= \frac{\pi}{\pi-2}\left(\frac{n}{n-1}\right)\sum_{i=1}^{n-1}\min(|\Delta p_i|,|\Delta p_{i+1}|)^2;
\]

\[
IV_{Med}= \frac{\pi}{6-4\sqrt{3}+\pi}\left(\frac{n}{n-2}\right)\sum_{i=2}^{n-1}\text{med}(|\Delta p_{i-1}|,|\Delta p_{i}|,|\Delta p_{i+1}|)^2.
\]

To make the estimators robust to the presence of microstructure noise, pre-averaging may be applied to price observations, as shown in \citet{andersen2012jump}, Appendix B. Accordingly, in this paper we used the noise-robust version of $IV_{Min}$ and $IV_{Med}$ with price pre-averaging.

\subsubsection*{Range estimator}
The main idea behind the range estimator by \citet{vortelinos2014optimally} is to substitute the simple returns with the difference between the maximum and minimum observed price over a given window, to obtain the estimator 

\[
IV_{RG} := \frac{1}{4\log(2)}\sum_{i=1}^{n/q}(max_{qi}-min_{qi})^2,
\]
where $max_{qi} = \max(p_{(i-1)q}, ..., p_{iq})$ and $min_{qi} = \min(p_{(i-1)q}, ..., p_{iq})$. The MSE-optimal frequency is 
\[
q^* =  \left( \frac{IQ}{\omega^4}\right)^{1/3}.
\]

\subsubsection*{Unified estimator}
 \citet{li2018unified} proposed a unified approach to volatility estimation, obtaining an estimator which is consistent not only in the presence of the typical i.i.d. noise, but also when the noise comes from price rounding. The estimators is defined as follows:
    
\[
IV_U:=\sum_{l=1}^m\left(\frac{1}{m}-\bar{n}\frac{n_l-\bar{n}}{\sum_{j=1}^Ln_j^2 -h\bar{n}^2}\right)\frac{1}{q_l}\sum_{k=0}^{q_l-1}\sum_{i=1}^{n_l} (p_{(k+i)q_l}-p_{k+(i-1)q_l})^2,
\]
where $\bar{n} = (\sum_{l=1}^m n_l)/h$, $n_l = n/q_l$ and  $q_{l+1} = q_1 +l$, $l =1,..., m-1$. The optimal $q_1$ and $h$ can be selected via the   data-driven procedure detailed in \citet{li2018unified}.

\subsection{Spot volatility}
 
\subsubsection*{Fourier estimator}
The Fourier method allows reconstructing the trajectory of the volatility as a function of time on a discrete grid, see \citet{mancino2015fourier}. This is achieved by means of the Fourier-Fej\'er inversion formula, which gives the estimator

$$\sigma^2_{F}(t) := \sum_{|k|\le M} \left(1-\frac{|k|}{M+1} \right)\textrm{e}^{\textrm{i}tk} c_k(\sigma^2_{n,N} ), \quad t \in [0,2\pi],$$

where 

$$c_k(\sigma^2_{n,N} )= \frac{2\pi}{2N+1}\sum_{|s|\le N}c_s(dp_n)c_{k-s}(dp_n)$$
 estimates  the k-th Fourier coefficient of the volatility.   Note that, differently from the other estimators detailed below, the Fourier estimator is $global$, in the sense that it estimates the entire volatility function on a discrete grid over the interval $[0,2\pi]$\footnote{By suitably re-scaling the unit of time, the estimator can be applied to any arbitrary interval.}, instead of a local value at a specific time $t$. 

The efficient selection of $N$ and $M$ can be performed based on the numerical results given \citet{mancino2015fourier}.

\subsubsection*{Regularized estimator}
Proposed by \citet{ogawa2008real}, the regularized estimator is based on a regularization procedure that involves data around the estimation point. The estimator reads

\[
\sigma_{REG}^2(t):= \frac{3s^2}{q\sqrt{[n/q]}(3sq-q^2+1)}\sum_{i=1}^{[n/q]} \Delta \bar{p}_{t+i}^2,
\]
where
\[
\bar{p}_t=\frac{1}{s}\sum_{j=1}^{s}p_{t q-j+1}.
\]

Following \citet{ogawa2011improved}, for the implementation we set $q=[n/n_t]$ and $s=2q$, where $n_t$ is the number of points on which the spot variance trajectory is reconstructed.

\subsubsection*{Kernel estimator}
\citet{fan2008spot} (see also \citet{kristensen2010nonparametric}) proposed an estimator of the spot variance based on the localization of the kernel-weighted realized variance over a window of length $q$, which reads

\[
\sigma_K^2(t):=\frac{1}{q}\sum_{t_i=t-q}^{t+q} k  \left(\frac{t_i -t}{q}\right)\Delta p_i^2,
\]
with
\[ q = [\sqrt{n}/ \log{n}].
\]

For the implementation, we used the Fejér kernel, following \citet{mancini2015spot}.

\subsubsection*{Pre-averaging estimator}
The pre-averaging estimator of the spot volatility by \citet{jing2014estimation} and \citet{li2021volatility}  relies on the localization of the pre-averaging integrated   estimator on a window of length $h$ and reads
    
\[
\sigma_{PA}^2(t) := \frac{n}{q}\frac{12}{s}\sum_{j=0}^{q-s+1}\frac{1}{2}\left(\frac{1}{s/2}\sum_{j'=0}^{s/2-1}p_{t+j+j'+s/2}
-\frac{1}{s/2}\sum_{j'=0}^{s/2-1}p_{t+j+j'}\right)^2-\frac{n}{q}\frac{6}{s}\sum_{j=0}^{q}\Delta p_{t+j}^2,
\]
with $q=[c_1n^{3/4}]$ and $s=[c_2n^{1/2}]$.
    
The constants $c_1$ and $c_2$  
can be chosen based on the numerical results in  \citet{li2021volatility}.

\subsubsection*{Two-scale estimator}
 The localized two-scale  estimator, proposed by \citet{zu2014estimating}, reads

\[
\sigma_{TS}^2(t):=\frac{1}{s}\sum_{j=q}^{n}\frac{(p_j-p_{j-q})^2}{q}  -\frac{\bar n}{n}\frac{1}{s}\sum_{j=1}^{n}   \Delta p_j^2,
\]
where  $\bar n=\frac{ns-q+1}{qs}$.
The MSE-{optimal} values of $s$ and $q$ are given by

\[ q^* = q^0n^{2/3} \qquad s^* = s^0n^{-1/6}, 
\]
where 
\[
q^0 = \left(\frac{12\omega^4}{IQ}\right)^{1/3} 
\]
and
\[
s^0 = \left( \frac{8\omega^4/(q^{0})^2+\frac{4}{3}q^0IQ}{\frac{1}{3}\gamma}\right)^{1/2},
\]
with
\[
\gamma =\sum_{i=1}^{n-1} (\Delta \sigma^2(t_i))^2, \quad \quad \sigma^2(t_i) = \frac{1}{s} \sum_{j=0}^{s-1} \Delta p_{t+j}^2.
\]

\subsubsection*{Optimal candlestick estimator}
The optimal candlestick estimator proposed by \citet{li2022reading} relies on candlestick data, i.e., the opening, closing, highest and lowest prices within a given interval. Formally, denote by $I_{n,i}:=\{ (i-1)\Delta_n, i\Delta_n\}$ the interval associated with the $i$th candlestick, where $\Delta_n \to 0$ as $n\to \infty$.    Moreover, let $O_i, C_i, H_i$ and $L_i$ denote, respectively, the opening, closing, highest and lowest prices on $I_{n,i}$. The estimator reads

$$\sigma^2_{OK}(t) = \frac{(\lambda_1 |O_i-C_i| + \lambda_2 (H_i-L_i))^2}{\Delta_n},   \quad t \in I_{n,i} $$

\noindent where $\lambda_1$ and $\lambda_2$ are constants. For the implementation, we set $\Delta_n=1$ minute (see Section \ref{sect:performanceCOMP}) and selected $(\lambda_1,\lambda_2)$ based on the optimality conditions detailed in \citet{li2022reading}, Section 2.2. 

\subsubsection*{Pre-averaging kernel estimator}
 \citet{figueroa2022kernel} proposed an estimator of the spot variance which consists in a localization of the realized kernel estimator with pre-averaging, i.e.,   
 \[
\sigma_{PAK}^2(t):=\frac{1}{f(g)}\sum_{j=1}^{n-s+1}\frac{1}{q}k\left(\frac{t_{j-1}-t}{q}\right)\left(\bar{p}_j^2-\frac{1}{2}\hat{p}_j\right)
\]

where:
\[
\bar{p}_j = -\sum_{i=1}^{s}\left(g\left(\frac{i}{s}\right)-g\left(\frac{i-1}{s}\right)\right)p_{i+j-2},
\]
\[
\hat{p}_j=\sum_{i=1}^{s}\left(g\left(\frac{i}{s}\right)-g\left(\frac{i-1}{s}\right)\right)^2\Delta p_{i+j-1}^2,
\]
    
$$f(g) =\sum_{i=1}^{s}g\left(\frac{i}{s}\right)^2,$$ 

\noindent with $g(x) = \min(x, (1-x))$ , $q=c_m(h/n)^{1/4}$ and $s= \lfloor 1/\left(c_k \sqrt{h/n} \right) \rfloor$.

The authors suggest to use the exponential kernel, that is,  $k(x) = \frac{1}{2} \exp(-|x|)$, as it is proved to be the optimal kernel in terms of the minimization of the asymptotic variance (see \citet{figueroa2020optimal}). Further, we choose $c_k$ and $c_m$ in accordance with the formulas derived by the authors to optimize the integrated asymptotic variance (see also Remark 4.1 in \citet{figueroa2022kernel}).

\subsection{Feasible selection of tuning parameters}

The feasible implementation of the optimization formulae for the tuning parameters which appear in the previous section may require the estimation of  $IV$, $IQ$ and $\omega^2$.  In this regard,  we use the following estimators, as in  \citet{gatheral2010zero}: 

\[
\widehat{IV} = \frac{n}{n-q+1}\frac{1}{q}\sum_{s=0}^{q-1}\sum_{i=1}^{\lfloor (n-s+1)/q \rfloor}(p_{iq+s}-p_{(i-1)q+s})^2
\]

\[
\widehat{IQ} =\left(\frac{n}{n-q+1}\right)^2\frac{26}{q}\sum_{s=0}^{q-1}\sum_{i=1}^{\lfloor (n-s+1)/q \rfloor}(p_{iq+s}-p_{(i-1)q+s})^4
\]
 
\[
\hat{\omega}^2 = -\hat{\phi}\hat{\sigma}^2_{w},  \,\,\,\, \hat{\omega}^4=(\hat{\omega}^2 )^2, 
\]
where the pair $(\hat \sigma^2_w, \hat \phi)$ is obtained as in (\ref{MLEnoise}).

Note that $q$ is selected in correspondence of the 5-minute (noise-free) sampling frequency.

\section{Comparative   performance study of volatility estimators}\label{sect:performanceCOMP}

In this section we present the results of a study of the finite-sample performance of the volatility estimators described in the previous section, which is based on  simulations of the QR model.  In the study, we considered two alternative scenarios.
In the first scenario, we assumed   constant values of the parameters $\theta$ and $\theta^{reinit}$, which translate  into a constant volatility parameter. Instead, in the second scenario we allowed   $\theta$ and $\theta^{reinit}$ to change, so that the volatility parameter is no longer constant. We emphasize the fact that this second scenario introduces a novelty compared to the study by \citet{gatheral2010zero}, where the volatility parameter is constant. In fact, a scenario with time-varying volatility offers a more realistic framework to assess the performance of estimators.  

 For each scenario, we simulated 2500 daily paths. For each couple $(\theta, \theta^{reinit})$ considered, the corresponding true value of the volatility parameter was obtained via additional simulations exploiting Eq. (\ref{truevol}), as described in Subsection \ref{features}.  Estimators were computed using 1-second price observations. The integrated volatility was estimated on daily intervals, while for the spot volatility we reconstructed daily trajectories on the 1-minute grid. The selection of the tuning parameters involved in the computation of the different estimators was performed based on the feasible formulae and the suggestions reported in the previous section.

 \subsection{Constant $\theta$ and $\theta^{reinit}$}
 
 In the first scenario, simulations were performed with  $\theta$ and $\theta^{reinit}$ constant and equal to, respectively, $0.6$ and $0.85$, that is, the parameter values calibrated on the MSFT sample (see Section \ref{sect:LOBmodels}). We recall that the resulting reference value of the spot variance parameter is $\sigma^2=1.0387 \cdot 10^{-8}$ (see Subsection \ref{features}).

Tables \ref{tab:intbias}-\ref{tab:spotMSE} display the ranking of the estimators for the three series of mid-price, micro-price and trade-price in terms of their finite-sample performance. The latter is evaluated by means of, resp.,  the relative bias and  MSE  for the integrated volatility and the relative integrated bias and  MSE for the spot volatility. In each table,   the estimators  are ordered based on the average ranking, obtained as the arithmetic mean of the rankings for the three prices series.

For what concerns integrated estimators, the pre-averaging estimator and the Fourier estimator provide  the relative best performance in terms, resp., of bias and MSE minimization, based on the average ranking. However, note that the Fourier estimator achieves the best MSE average ranking without resulting the first in the ranking for any price series. 
\normalcolor 
Indeed, the unified volatility estimator, which is robust to i.i.d. noise and rounding, yields the best performance for mid- and micro-prices, while the best result for trade-prices is achieved by the alternation estimator, which is robust to price  discreteness and rounding  (see \citet{large2011estimating}.  Overall, this  may suggest that robustness to rounding may be crucial to optimize the mean squared error. Instead, the pre-averaging estimator is more clearly favorable in terms of bias reduction, given the fact that it results the first in terms of bias minimization for  micro- and trade-price series and the second for  mid-price series. Moreover, it is worth noticing that the Min RV and Med RV, which are also computed from pre-averaged data, occupy the second and third places in the average ranking for the bias. 
\normalcolor

\begin{figure}[htbp!]
    \centering
    \includegraphics[width =1\linewidth]{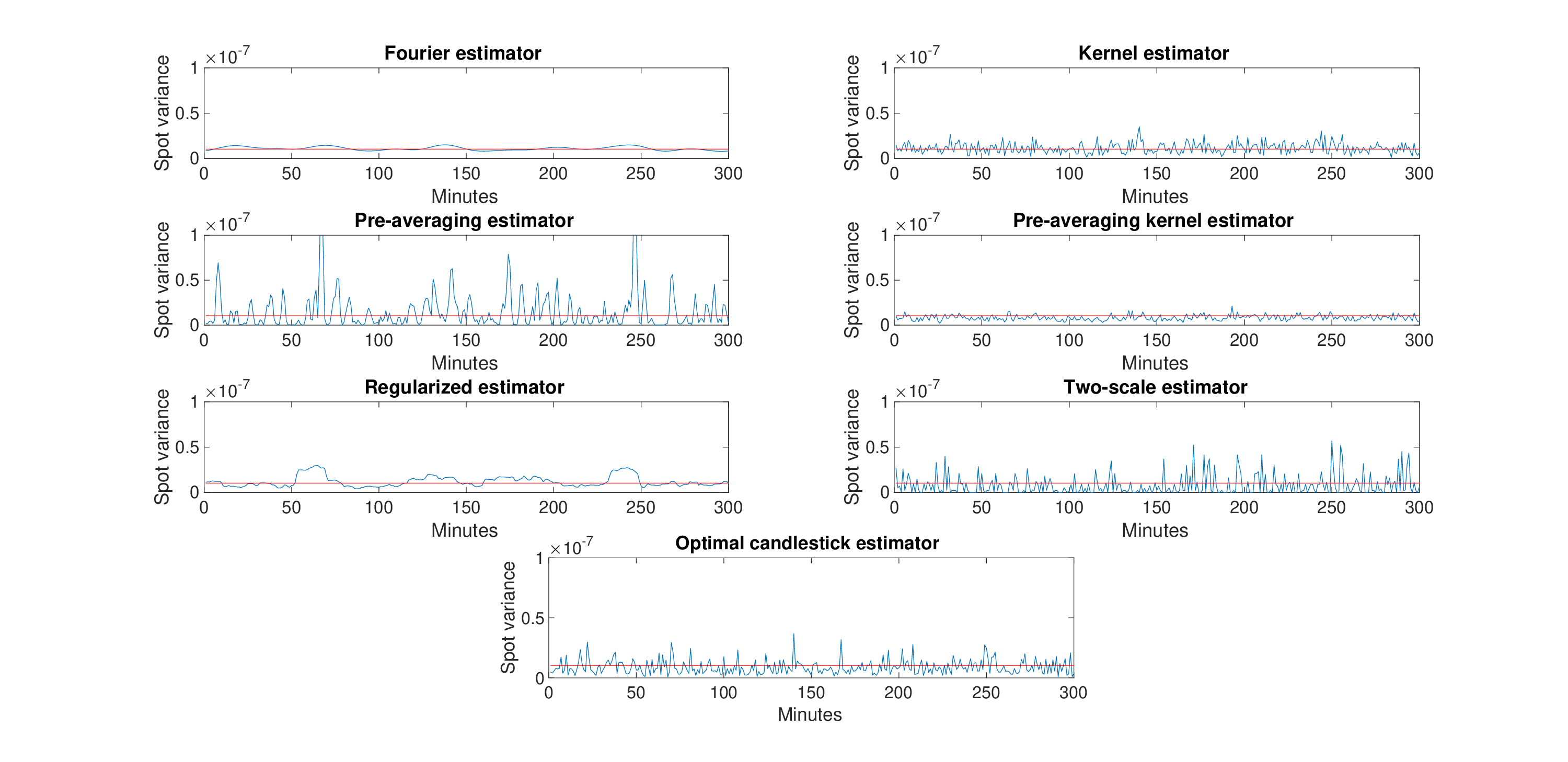}
    \caption{Constant $\theta$ and $\theta^{\text{reinit}}$: sample trajectories of spot variance estimators computed from mid-prices (in blue) and true volatility (in red).}
    \label{fig:spotmid}
\end{figure}

As for spot estimators,  simulations indicate that the Fourier estimator outperforms the other estimators considered, both in terms of bias and MSE optimization, for all the price series considered. Only the regularized estimator is capable of obtaining comparable performances in terms of bias. The Fourier estimator differs from the other spot estimators considered in that  it relies on the integration of the Fourier coefficients of the volatility rather than on the differentiation of the (estimated) integrated variance, and this appears to represent a solid numerical advantage.  Figure \ref{fig:spotmid} shows sample trajectories of the spot estimators computed from the mid-price series, along with the true volatility value, to help better understand the difference in performance among the estimators.   Appendix A.2 contains the analogous figures for the micro-price and the trade-price.

Finally, in the case of both integrated and spot estimators we investigated whether the volatility values obtained using different methods were statistically different. To this end, for each pair of estimators, we performed a t-test of the null hypothesis that the mean value of the estimated volatility is the same. As a result, we found that all   average estimations are  pairwise significantly different  at the 1\% confidence level. Moreover, to better assess the differences in the performances of estimators,  we also applied the model confidence set (MCS) procedure by \citet{hansen2011model}, with a significance level of 1\%. Following \citet{patton2011volatility}, for the MCS we used the qlike loss function. 
As a result, the MCS procedure always chooses as the optimal model set the one containing only the estimator with the best ranking in terms of MSE (see Tables \ref{tab:intMSE} and \ref{tab:spotMSE}), thus supporting the soundness of our results.
Overall, the results of the t-tests and the MCS offer additional support to the fact  that the careful selection  of the estimation method is not  irrelevant.

\begin{table}[htbp!]
\begin{center}
    \begin{tabular}{c|c|c|c|c|c|c|c}
        \hline\hline
        \multicolumn{8}{c}{Integrated variance estimators - relative bias}\\
		\hline\hline
		Estimator  & mid-price & rank & micro-price & rank & trade-price & rank &
		av. rank\\
		\hline
		Pre-averaging & -0.01106 &2  & -0.00643 &1 & -0.00534 & 1 & 1.33 \\
		Min RV & 0.01071 & 1  & 0.01050 & 2 & 0.01181 & 3 & 2 \\
		Med RV & 0.01367  &3  & 0.01348 & 3 & 0.01456 & 4 & 3.33 \\
		Unified  & 0.05518 & 4  & 0.04679 & 4 & 0.11442 & 7 & 5\\
		Fourier & 0.09154 & 6 & 0.08290 & 5 & 0.09234 & 6 & 5.66 \\
		Range & -0.09196 & 7 & -0.08302 & 6 & -0.06997 &  5& 6\\
		Alternation & 0.06799 & 5  & 0.11612 & 12 & -0.06226 & 2 & 6.33\\
		Kernel & 0.12679 & 8 & 0.10075 & 8 & 0.24836 & 8 & 8 \\
		Maximum likelihood & 0.13990 & 12 & 0.10011 & 7 & 0.27243 & 9 & 9.33 \\
		Two-scale RV & 0.12827 & 9 & 0.10465 & 9  & 0.32015 & 10 & 9.66 \\
		Multi-scale RV & 0.12841 & 10 & 0.10478 & 10 & 0.32030 & 11 & 10.33 \\
		Bias-corrected RV & 0.12852 & 11  & 0.10489 &11 & 0.32042 & 12 &11.33 
    \end{tabular}
    \caption{Performance and ranking of the integrated variance estimators for the series of mid-price, micro-price and trade-price, according to the relative bias. The average ranking is reported in the last column.}
	\label{tab:intbias}
\end{center}
\end{table}

\begin{table}[htbp!]
\begin{center}
    \begin{tabular}{c|c|c|c|c|c|c|c}
        \hline\hline
        \multicolumn{8}{c}{Integrated variance estimators - relative MSE}\\
		\hline\hline
		Estimator & mid-price & rank & micro-price & rank & trade-price & rank &
		av. rank\\
		\hline
		Fourier & 0.00981 & 3 & 0.00658 &2 & 0.01039 & 2 & 2.33 \\
		Unified  & 0.00433 & 1  & 0.00346 & 1 & 0.01446 & 6 &2.66\\
		Alternation & 0.00593 & 2  & 0.01145  & 5 & 0.00485 &1 & 2.66\\
		Med RV & 0.01075 & 4 & 0.01174 & 6 & 0.01181 & 3 & 4.33\\
		Min RV & 0.01275 & 5 & 0.012754 & 7 & 0.01282 & 4 & 5.33\\
		Maximum likelihood & 0.01647  & 7 & 0.01112 & 3 & 0.07564 & 9 & 6.33\\
		Kernel & 0.01726 & 9 & 0.01128 & 4 & 0.06314 & 8 & 7\\
		Pre-averaging & 0.01566 & 6 & 0.015653 &11 & 0.01568 & 7 & 8\\
		Multi-scale RV & 0.01648 & 8 & 0.012070 & 8 & 0.10401 & 11 & 9\\
		Range & 0.01794  & 12 & 0.016448 & 12 & 0.01304 & 5 & 9.66\\
		Two-scale RV & 0.01757  & 10 & 0.012901 & 10 & 0.10391 & 10 & 10\\
		Bias-corrected RV & 0.01734  &  11 & 0.012094  & 9 & 0.10409 &12 & 10.66
    \end{tabular}
    \caption{Performance and ranking of the integrated variance estimators for the series of mid-price, micro-price and trade-price, according to the relative mean squared error. The average ranking is reported in the last column.}
	\label{tab:intMSE}
\end{center}
\end{table}

\begin{table}[htbp!]
\begin{center}
    \begin{tabular}{c|c|c|c|c|c|c|c}
        \hline\hline
        \multicolumn{8}{c}{Spot variance estimators - relative integrated bias}\\
		\hline\hline
		Estimator & mid-price & rank & micro-price & rank & trade-price & rank &
		av. rank\\
		\hline
        Fourier & 0.00362 & 1 & 0.00299 & 1& 0.00713 & 1 & 1\\
        Regularized & 0.00676 & 2 & 0.00663 &2 & 0.00802 & 2 & 2\\
        Optimal candlestick & -0.16165 & 3 & -0.12525 & 3 & -0.07430 &  3  & 3\\
        Pre-averaging kernel & -0.14755 &4 & -0.13462 & 4 & -0.08459 & 4 & 4\\
        Pre-averaging & 0.19939 & 5 & 0.19923 & 5 & 0.19936 & 6 & 5.33\\
        Kernel & 0.22205 & 6 & 0.23801 & 6 & 0.73200 & 7 & 6.33 \\
        Two-scale & -0.24781  & 7 & -0.26271 & 7 & -0.11798 &  5 & 6.33 \\
    \end{tabular}
    \caption{Performance and ranking of the spot variance estimators for the series of mid-price, micro-price and trade-price, according to the relative integrated bias. The average ranking is reported in the last column.}
	\label{tab:spotbias}
\end{center}
\end{table}

\begin{table}[htbp!]
\begin{center}
    \begin{tabular}{c|c|c|c|c|c|c|c}
        \hline\hline
        \multicolumn{8}{c}{Spot variance estimators - relative integrated MSE}\\
		\hline\hline
		Estimator & mid-price & rank & micro-price & rank & trade-price & rank &
		av. rank\\
		\hline
        Fourier & 0.03095  & 1 & 0.03060 & 1 & 0.03128 & 1&1 \\
        Pre-averaging kernel & 0.13392 & 2 & 0.13745 & 2 & 0.09148 &2 &2 \\
        Regularized & 0.20632 & 3 & 0.20628 & 3 & 0.20635 &3 &3 \\
        Optimal candlestick & 0.31790 & 4 & 0.31436 & 4 & 0.33429 & 4&4 \\
        Kernel & 0.46111 & 5 & 0.47095 & 5 & 0.84365 & 5&5 \\
        Two-scale & 1.62190  & 6 & 1.59843 & 6 & 2.24599 & 6&6 \\
        Pre-averaging & 2.59134 & 7 & 2.59064 & 7 & 2.59247 & 6&7 \\
    \end{tabular}
    \caption{Performance and ranking of the spot variance estimators for the series of mid-price, micro-price and trade-price, according to the relative integrated mean squared error. The average ranking is reported in the last column.}
	\label{tab:spotMSE}
\end{center}
\end{table}

\pagebreak

\subsection{Variable \texorpdfstring{$\theta$ and $\theta^{\text{reinit}}$}{theta and thetareinit}}

A nice feature of the QR model, compared to the ZI model, is the flexibility introduced by the parameters $\theta$ and $\theta^{\text{reinit}}$. In the second scenario we assessed the effect  of   time-varying values of $\theta$ and $\theta^{\text{reinit}}$ on the accuracy of volatility estimators. Specifically, we allowed for piece-wise constant volatility dynamics, which might describe  a regime-shifting scenario driven, for example, by the flow of information hitting the market.

We considered two sub-scenarios, with increasing variability of the volatility parameter. In the first one, the LOB follows five regimes (each with length equal to $1/5$ of a day), which translate into a double u-shaped pattern for the volatility parameter, as illustrated in Table \ref{tab:5lv}.

\begin{table}[htbp!]
	\begin{center}
		\begin{tabular}{c|c|c|c}
			\hline\hline
			 Period & $\theta$ & $\theta^{\text{reinit}}$ & $\sigma^2 $\\
			\hline\hline
			I & 0.7 & 0.6 & $1.473\cdot 10^{-8}$ \\
			II & 0.4 & 0.6 & $8.553 \cdot 10^{-9}$\\
			III & 0.6 & 0.85 & $1.039\cdot 10^{-8}$\\
			IV & 0.4 & 0.9 & $7.195\cdot 10^{-9}$ \\
			V & 0.8 & 0.9 & $1.366\cdot 10^{-8}$
		\end{tabular}
		\caption{$\theta$ and $\theta^{\text{reinit}}$ values used for the simulations with five regimes and the corresponding values of $\sigma^2$.}
		\label{tab:5lv}
	\end{center}
\end{table}

In the second sub-scenario, we allowed for ten regimes (each with length equal to $1/10$ of a day), which recreate  a u-shape pattern for the volatility parameter, see Table \ref{tab:10lv}.

  \begin{table}[htbp!]
	\begin{center}
		\begin{tabular}{c|c|c|c|c|c|c|c}
			\hline\hline
			 Period & $\theta$ & $\theta^{\text{reinit}}$ & $\sigma^2 $& Period & $\theta$ & $\theta^{\text{reinit}}$ & $\sigma^2 $\\
			\hline\hline
			I & 0.7 & 0.6 & $1.473\cdot 10^{-8}$ &
			VI & 0.3 & 0.5 & $4.813\cdot 10^{-9}$\\
			II & 0.8 & 0.9 & $1.366\cdot 10^{-8}$ &
			VII & 0.4 & 0.9 & $7.195\cdot 10^{-9}$\\
			III & 0.8 & 0.7 & $1.281\cdot 10^{-8}$ &
			VIII & 0.4 & 0.6 & $8.553 \cdot 10^{-9}$ \\
			IV & 0.5 & 0.5 & $7.753\cdot 10^{-9}$ &
			IX & 0.6 & 0.85 & $1.039\cdot 10^{-8}$\\
			V & 0.2 & 0.8 & $3.068\cdot 10^{-9}$ &
			X & 0.9 & 0.4 & $1.469\cdot 10^{-8}$\\
		\end{tabular}
		\caption{$\theta$ and $\theta^{\text{reinit}}$ values used for the simulations with ten regimes and the corresponding values of $\sigma^2$.}
		\label{tab:10lv}
	\end{center}
\end{table}

Tables \ref{tab:complvint} and \ref{tab:complvspot} illustrate  the average performance rankings in the second scenario. Specifically, average rankings in correspondence of five and ten intra-day regimes are compared with the case of a unique regime (i.e., the case with constant $\theta$ and $\theta^{reinit}$ illustrated in the previous subsection). Full performance results in terms of bias and MSE are detailed in Appendix A.1.

Overall,  it appears that the introduction of time-varying parameters does not significantly affect the performance rankings previously obtained with constant parameters. In other words,  most of the rankings are quite stable (with a few exceptions, see, e.g., the average ranking for the MSE of the Range  and Pre-averaging estimators), compared to the first scenario. The stability is more evident  for spot variance estimators. 

As a further investigation, it might be interesting to study whether the performance ranking remains similar also when the volatility path has infinite regimes, for example when the evolution of $\theta$ and $\theta^{reinit}$ is driven by stochastic differential equations. This would require to build a precise mapping between the  volatility and $\theta$ and $\theta^{reinit}$ in order to simulate the LOB for each value of the simulated volatility. This interesting investigation is beyond the scope of this paper and is left for future research.

\begin{figure}[htbp!]
    \centering
    \includegraphics[width =1\linewidth]{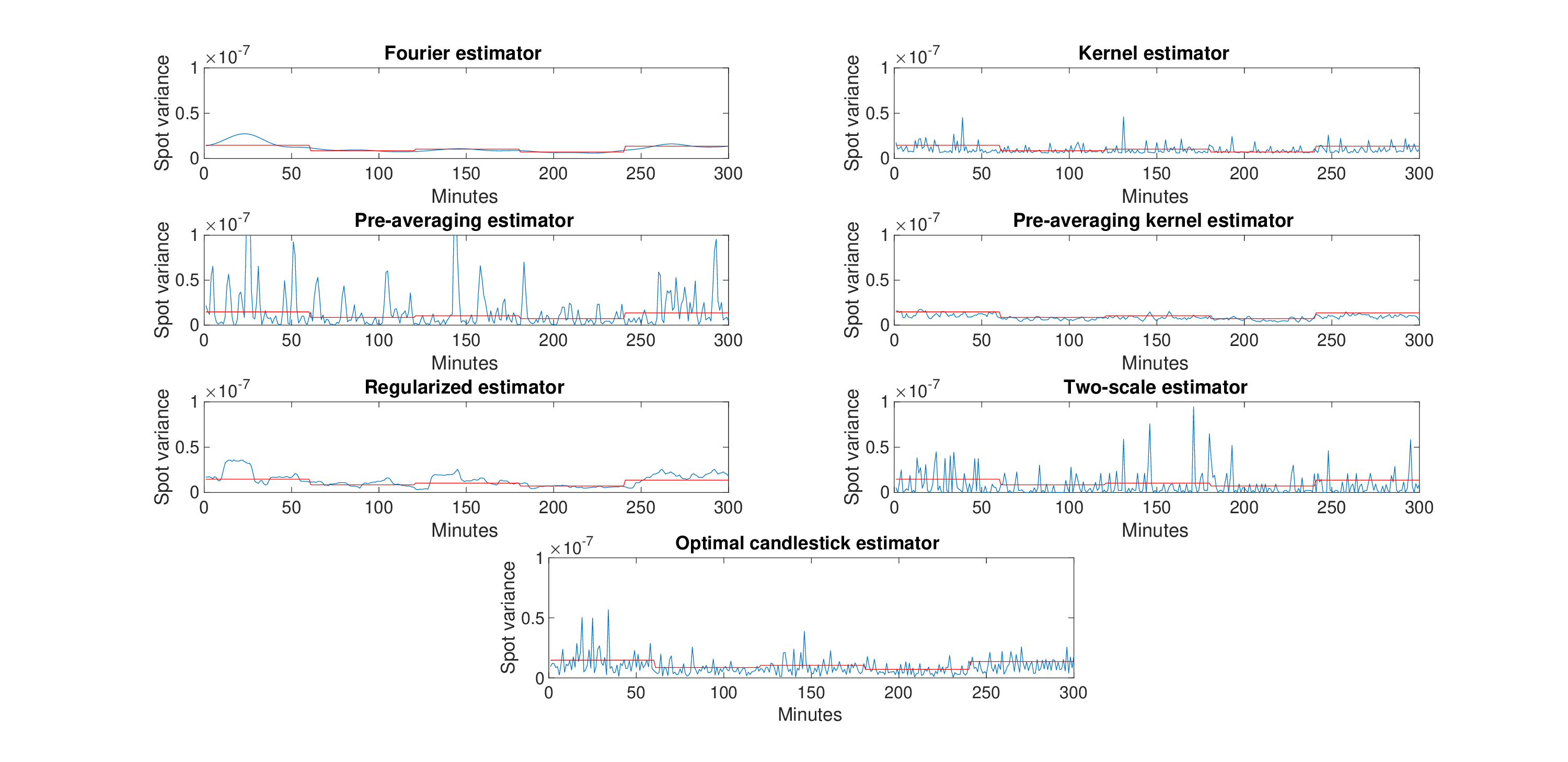}
    \caption{Variable $\theta$ and $\theta^{\text{reinit}}$ (5 regimes): sample trajectories of spot variance estimators computed from mid-prices (in blue) and true volatility (in red).}
    \label{fig:spotmid5lv}
\end{figure}

\begin{figure}[htbp!]
    \centering
    \includegraphics[width =1\linewidth]{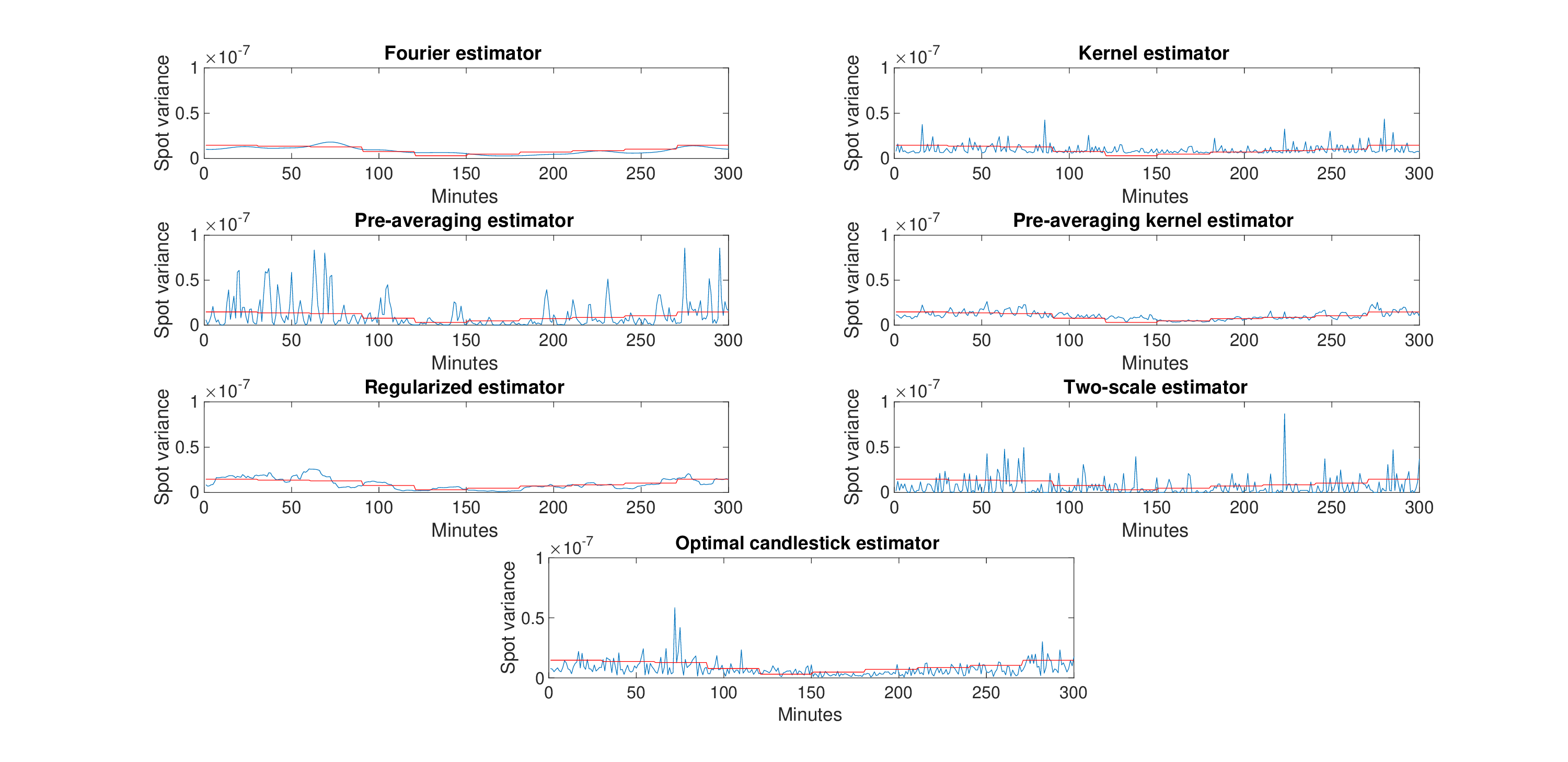}
    \caption{Variable $\theta$ and $\theta^{\text{reinit}}$ (10 regimes): sample trajectories of spot variance estimators computed from mid-prices (in blue) and true volatility (in red).}
    \label{fig:spotmid10lv}
\end{figure}

As in the previous subsection, to help better understand the difference in performance among the estimators in the second scenario, Figures \ref{fig:spotmid5lv}  and \ref{fig:spotmid10lv} contain sample trajectories of the spot variance estimators computed from  mid-price observations, together with the path of the true  variance parameter; the analogous figures for   micro- and trade-prices are in Appendix A.2.
  
\normalcolor

\begin{table}[htbp!]
    \centering
    \begin{tabular}{c|c|c|c|c|c|c}
    \hline\hline
    \multicolumn{7}{c}{Average rankings for different volatility regimes (integrated variance)}\\
    \hline\hline
     & \multicolumn{3}{c}{relative bias} & \multicolumn{3}{|c}{relative MSE}\\
    \hline
     Estimator & 1 regime & 5 regimes &10  regimes & 1 regime  & 5 regimes &10 regimes \\
     \hline	
       Pre-averaging  & 1.33 & 1 &2& 8 & 4 & 4.33\\
       Fourier  & 5.66 &6.33 &6.33& 2.33 & 3 & 3.33\\
       Med RV  & 3.33 &2 &1& 4.33 &5.33 & 6.66\\
		Min RV  & 2 & 3& 3.33& 5.33 & 7 & 8.66\\
        Unified   & 5 & 6& 5.66& 2.66 & 3 & 3.33\\
		
	    Range  & 6 & 4 &5 & 9.66 & 3 & 3.33\\
		Alternation  & 6.33 & 7.66 &6.33& 2.66 & 5 & 4.33\\
		
		Maximum likelihood  & 9.33 &8.66 & 8& 6.33 &7.66 & 6.33\\
		Kernel  &8 & 8.33 & 9& 7 & 8 & 8\\
		Two-scale RV  & 9.66 &9.33 & 9.33& 10 & 9.66 & 9.66\\
		Multi-scale RV  & 10.33 & 10.33 & 10.66& 9 & 10.66 & 9\\
		Bias-corrected RV  &11.33 & 11.66 & 11.66& 10.66 & 11.66 & 11\\ 
    \end{tabular}
    \caption{Average rankings of integrated variance estimators for an increasing number of intra-day volatility regimes.}
    \label{tab:complvint}
\end{table}

\begin{table}[htbp!]
    \centering
    \begin{tabular}{c|c|c|c|c|c|c}
    \hline\hline
    \multicolumn{7}{c}{Average rankings for different volatility regimes (spot variance)}\\
    \hline\hline
     & \multicolumn{3}{c}{relative int. bias} & \multicolumn{3}{|c}{relative int. MSE}\\
    \hline
     Estimator & 1 regime  & 5 regimes &10 regimes & 1 regime  & 5 regimes &10 regimes\\
     \hline
         Fourier & 1 & 1 & 1 & 1 & 1 & 1\\
          Regularized  & 2 &2 & 2& 3 & 3 & 3\\
          Pre-averaging kernel  & 4 &3 & 3 & 2 & 2 & 2\\
        
         Optimal candlestick  & 3 & 4 & 4.33& 4 & 4 & 4\\
         Kernel &  6.33 & 6 & 5.66& 5 & 5 & 5  \\
         Pre-averaging  & 5.33 & 5 & 5.66& 7 & 7 & 6.66\\
         Two-scale  & 6.33 & 7 & 7& 6 & 6 & 6.33\\
    \end{tabular}
    \caption{Average rankings of spot variance estimators for an increasing number of intra-day volatility regimes.}
    \label{tab:complvspot}
\end{table}

 \newpage
\section{The impact of efficient volatility estimates on optimal execution}\label{sec:optex}

 {In this section, we discuss the results of a study which aims at providing insights into the impact of the use of efficient volatility estimates on the prediction of the variance of the cost of a VWAP execution.}

Consider the following problem. A trader has $ S $ shares to buy within the interval $[0,T]$. The interval is divided into $N_{\tau}$ time periods of length $\tau=T/N_{\tau}$ and $v_k$, $k=1,...,N_{\tau}$, denotes the (signed) number of shares to be traded in interval $k$. Clearly, $\sum_{k=1}^{N_{\tau}}v_k=S$. Moreover, let $\tilde p_k$ be the price at which the investor trades in interval $k$ (in general different from the average price in the interval, $p_k$) and $p_0$ the price before the start of the execution.
The objective function is the {\it Implementation Shortfall} (IS), defined as
\begin{equation}\label{eq:empcost}
C(\mathbf{v})\equiv \sum_{k=1}^{N_{\tau}} v_k\tilde p_k-Sp_0
\end{equation}
i.e., as the difference between the cost and the cost in an infinitely liquid market. The IS is in general a stochastic variable, therefore one often wishes to minimize 
$$
{\mathbb E}[ C(\mathbf{v})]+\lambda Var[C(\mathbf{v})]
$$
where $\lambda$ measures the risk aversion of the trader. 

Let us consider a traders which models markt impact according to the Almgren and Chriss model (\citet{almgren2001optimal}). In this model, the price of the stock at step $k$ is equal to the previous price plus a linear permanent market impact term and a random shock, that is,
\begin{equation}
p_k = p_{k-1}  + \vartheta v_{k} + u_{k}\;\;\;\;~~ u_k \sim \mbox{IID}(0, \sigma^2 \tau),
\label{priceDyn}
\end{equation}
where $\sigma^2$ is instantaneous volatility of the unaffected price.
Moreover, the actual price paid $\tilde{p}_k$  is different from the average price $p_k$ in the interval and reads
\begin{equation}
  \tilde{p}_k = p_{k}  + \rho  v_{k}, 
  \label{eq:almgren_chriss_effective_price}
\end{equation}
where $ \rho v_{k}$ represents a linear temporary impact.

The expected cost of an execution is then given by
$$
E[C(\mathbf{v})]=(\vartheta+\rho) \sum_{k=1}^{N_{\tau}} v_k^2+\vartheta \sum_{i>j}v_iv_j
$$
and its variance is equal to 
$$
Var[C(\mathbf{v})]= \sigma^2 \tau \mathbf{v}^T B \mathbf{v},
$$
where  
$$
B=\left(\begin{array}{ccccc}1 & 1 & 1 & 1 & ... \\1 & 2 & 2 & 2 & ... \\1 & 2 & 3 & 3 & ... \\1 & 2 & 3 & 4 & ... \\... & ... & ... & ... & ...\end{array}\right).
$$
Note that the variance {does not} depend on the impact parameters $\rho$ and $\theta$, but only on the volatility. Further, note that the above expression is  more general, as it remains valid also when the temporary impact is nonlinear, see \citet{gueant2016financial} for more details.

For the sake of simplicity we assume that the trader performs a VWAP execution (i.e., $\mathbf{v}=\frac{S}{N_{\tau}}(1,1,...,1)^T$), so that the variance of the cost is equal to
\begin{equation}\label{eq:varvwap}
Var[C_{VWAP}]= \frac{\sigma^2S^2}{N_{\tau}^2} \tau\sum_{k=0}^{N_{\tau}-1}(2k+1)(N_{\tau}-k)=S^2\sigma^2\tau \frac{1}{6N_{\tau}}(2N_{\tau}^2+3N_{\tau}+1)
\end{equation}

This expression shows that, in order to estimate the variance of the execution cost, the traders must have a reliable estimation of $\sigma$. We  investigate whether the availability of an efficient estimate of the latent volatility parameter could allow the trader to reliably infer the variance of the cost of the strategy. More specifically, we are interested in assessing whether the use of a specific spot volatility estimator, among those studied in Section \ref{sect:performanceCOMP}, leads to a gain in the accuracy of such inference.

To this aim, we use Monte Carlo scenarios of the QR model to simulate a VWAP execution and we compare the variance of the cost of the simulated executions  with the corresponding value predicted by the Almgren and Chriss model (see Eq. (\ref{eq:varvwap})), evaluated with the (average) value of $\sigma^2$ obtained through a specific estimator. 
Since it is not obvious that the Almgren-Chriss model faithfully describes the market impact in the QR model,
 we opted for a more robust comparison that considers the ratio of the aforementioned quantities in correspondence of two different values of the couple $(\theta, \theta^{reinit})$, i.e. of the volatility.  In this way, the effect of the strategy parameters $S, \tau$ and $N_\tau$ in equation \ref{eq:empcost}, that depend on the specific market-impact model, disappear.

For the simulation of the execution strategy, we set   $T=$ $3$ hours and $20$ minutes, $\tau=10$ minutes and $S=60$, so that  $N_{\tau}= 20$ and  $\textbf{v}=(3,...,3)^T$. The strategy was simulated on top of   QR dynamics simulated in two different scenarios with parameters $(\theta, \theta^{reinit})$ equal to $(0.6,0.85)$ and $(0.4,0.6)$. We considered $100$ VWAP executions and we computed the empirical variance of their execution cost. The average spot variance values were retrieved from the study of Section \ref{sect:performanceCOMP}.

 Table \ref{tab:vwapcostvar} compares the ratios  obtained for each spot variance estimator with the benchmark ratio, that is, the ratio of empirical variance costs. Values related to   $(\theta, \theta^{reinit})=(0.6,0.85)$ (respectively,  $(\theta, \theta^{reinit})=(0.4,0.6)$) were used at the numerator (respectively, denominator).

\begin{table}[htbp]
	\begin{center}
		\begin{tabular}{c|c}
			\hline\hline
			\multicolumn{2}{c}{Ratios of variances  of the VWAP execution cost}\\
			\hline\hline
			Empirical variance & 1.397 \\
			\hline
			Fourier  & 1.235  \\
			Regularized    & 1.234 \\
			Pre-averaging kernel    & 1.563 \\
			Optimal candlestick & 1.566 \\
			Pre-averaging    & 1.208 \\
			Kernel    & 1.186 \\
			Two-scale    &  1.030  \\
		\end{tabular}
		\caption{Ratios of variances  of the VWAP implementation shortfall in correspondence of, respectively, $(\theta, \theta^{reinit})=(0.6,0.85)$ (numerator) and  $(\theta, \theta^{reinit})=(0.4,0.6)$ (denominator).}
		\label{tab:vwapcostvar}
	\end{center}
\end{table}

 Results in Table \ref{tab:vwapcostvar} suggest that the Fourier  estimator and  the regularized estimator    produce the relative best forecasts of the variance of the strategy costs, as they are associated to a ratio approximately equal to 1.23, which is the closest to the benchmark value of 1.397. As these two estimators provide also the relative best performance in terms of bias and MSE (see Section \ref{sect:performanceCOMP}), our study    suggests  that efficient volatility estimates  may be linked to a better forecast of the variance of the execution cost. Furthermore, note that the range of variation of the ratios in Table \ref{tab:vwapcostvar} suggests that the the choice of the estimator is not irrelevant and may lead to significant differences in the forecast of the execution strategy. It seems however that, in general, the use of the formula in Eq. (\ref{eq:varvwap}) leads to a certain underestimation of the the variance of the implementation shortfall of the considered strategy.

\section{Conclusions}\label{sect:conclusions}

This paper extended the  work by \citet{gatheral2010zero} on volatility estimation with LOB data in two directions.   First, we used  a more sophisticated LOB simulator compared to the ZI model, namely the QR model, which, by introducing correlations between the current state of the LOB and the intensities of order arrival, and thanks to a more complex re-initialization mechanism, is able to produce   more realistic market microstructure dynamics.

Secondly, we addressed not only integrated volatility estimators, but also spot volatility estimators. For what concerns   integrated estimators, we found that   the pre-averaging estimator by \citet{jacod2009microstructure} appears to be favorable in terms of   bias optimization. Instead, when looking at the minimization of the MSE, the situation is more nuanced, with the   Fourier estimator by \citet{malliavin2009fourier} obtaining the best average ranking across the three different price series considered without actually achieving the best    ranking for any of the individual series. Specifically, the  MSE is optimized by the  unified estimator by \citet{li2018unified} (in the case of mid- and micro-prices) and the alternation estimator by \citet{large2011estimating} (in the case of trade-prices). As for the spot volatility,   the Fourier estimator   appeared to yield   the  optimal accuracy both in terms of bias and MSE, outperforming the other estimators considered.  

 Finally, our results suggested that the careful choice of the spot volatility estimator may be relevant for optimal execution. Specifically, we investigated the impact of   different spot  volatility  estimators on  the prediction of the variance of the cost of  a VWAP strategy and found   that the use of the Fourier estimator, which  gave the  relative most accurate volatility estimates,   lead  also to a significant gain in   predicting   the cost variance. 
 
\section*{Acknowledgments}

The authors are thankful to  Jim Gatheral and Mathieu Rosenbaum for the useful discussions and to Othmane Mounjid for the valuable insights into the implementation of the Queue-Reactive model.

\printbibliography

@book{abergel2016limit,
  title={Limit order books},
  author={Abergel, Fr{\'e}d{\'e}ric and Anane, Marouane and Chakraborti, Anirban and Jedidi, Aymen and Toke, Ioane Muni},
  year={2016},
  publisher={Cambridge University Press}
}

@book{ait2014high,
  title={High-frequency financial econometrics},
  author={A{\"i}t-Sahalia, Yacine and Jacod, Jean},
  year={2014},
  publisher={Princeton University Press}
}

@article{ait2005often,
  title={How often to sample a continuous-time process in the presence of market microstructure noise},
  author={A{\"i}t-Sahalia, Yacine and Mykland, Per A and Zhang, Lan},
  journal={The review of financial studies},
  volume={18},
  number={2},
  pages={351--416},
  year={2005},
  publisher={Oxford University Press}
}

@article{ait2019hausman,
  title={A Hausman test for the presence of market microstructure noise in high frequency data},
  author={A{\"i}t-Sahalia, Yacine and Xiu, Dacheng},
  journal={Journal of Econometrics},
  volume={211},
  number={1},
  pages={176--205},
  year={2019},
  publisher={Elsevier}
}

@article{almgren2001optimal,
  title={Optimal execution of portfolio transactions},
  author={Almgren, Robert and Chriss, Neil},
  journal={Journal of Risk},
  volume={3},
  pages={5--40},
  year={2001}
}

@article{andersen2012jump,
  title={Jump-robust volatility estimation using nearest neighbor truncation},
  author={Andersen, Torben G and Dobrev, Dobrislav and Schaumburg, Ernst},
  journal={Journal of Econometrics},
  volume={169},
  number={1},
  pages={75--93},
  year={2012},
  publisher={Elsevier}
}

@article{barndorff2008designing,
  title={Designing realized kernels to measure the ex post variation of equity prices in the presence of noise},
  author={Barndorff-Nielsen, Ole E and Hansen, Peter Reinhard and Lunde, Asger and Shephard, Neil},
  journal={Econometrica},
  volume={76},
  number={6},
  pages={1481--1536},
  year={2008},
  publisher={Wiley Online Library}
}

@book{bouchaud2018trades,
  title={Trades, quotes and prices: financial markets under the microscope},
  author={Bouchaud, Jean-Philippe and Bonart, Julius and Donier, Jonathan and Gould, Martin},
  year={2018},
  publisher={Cambridge University Press}
}

@article{delbaen1994general,
  title={A general version of the fundamental theorem of asset pricing},
  author={Delbaen, Freddy and Schachermayer, Walter},
  journal={Mathematische annalen},
  volume={300},
  number={1},
  pages={463--520},
  year={1994}
}

@article{fan2008spot,
  title={Spot volatility estimation for high-frequency data},
  author={Fan, Jianqing and Wang, Yazhen},
  journal={Statistics and its Interface},
  volume={1},
  number={2},
  pages={279--288},
  year={2008},
  publisher={International Press of Boston}
}

@article{figueroa2020optimal,
  title={Optimal kernel estimation of spot volatility of stochastic differential equations},
  author={Figueroa-L{\'o}pez, Jos{\'e} E and Li, Cheng},
  journal={Stochastic Processes and their Applications},
  volume={130},
  number={8},
  pages={4693--4720},
  year={2020},
  publisher={Elsevier}
}

@article{figueroa2022kernel,
  title={Kernel estimation of spot volatility with microstructure noise using pre-averaging},
  author={Figueroa-L{\'o}pez, Jos{\'e} E and Wu, Bei},
  journal={arXiv preprint arXiv:2004.01865},
  year={2022}
}

@article{gatheral2010zero,
  title={Zero-intelligence realized variance estimation},
  author={Gatheral, Jim and Oomen, Roel CA},
  journal={Finance and Stochastics},
  volume={14},
  number={2},
  pages={249--283},
  year={2010},
  publisher={Springer}
}

@book{gueant2016financial,
  title={The Financial Mathematics of Market Liquidity: From optimal execution to market making},
  author={Gu{\'e}ant, Olivier},
  volume={33},
  year={2016},
  publisher={CRC Press}
}

@article{hansen2006realized,
  title={Realized variance and market microstructure noise},
  author={Hansen, Peter R and Lunde, Asger},
  journal={Journal of Business \& Economic Statistics},
  volume={24},
  number={2},
  pages={127--161},
  year={2006},
  publisher={Taylor \& Francis}
}

@article{hansen2011model,
  title={The model confidence set},
  author={Hansen, Peter R and Lunde, Asger and Nason, James M},
  journal={Econometrica},
  volume={79},
  number={2},
  pages={453--497},
  year={2011},
  publisher={Wiley Online Library}
}

@book{hasbrouck2007empirical,
  title={Empirical market microstructure: the institutions, economics, and econometrics of securities trading},
  author={Hasbrouck, Joel},
  year={2007},
  publisher={Oxford University Press}
}

@article{huang2015simulating,
  title={Simulating and analyzing order book data: The queue-reactive model},
  author={Huang, Weibing and Lehalle, Charles-Albert and Rosenbaum, Mathieu},
  journal={Journal of the American Statistical Association},
  volume={110},
  number={509},
  pages={107--122},
  year={2015},
  publisher={Taylor \& Francis}
}

@article{jacod2009microstructure,
  title={Microstructure noise in the continuous case: The pre-averaging approach},
  author={Jacod, Jean and Li, Yingying and Mykland, Per A and Podolskij, Mark and Vetter, Mathias},
  journal={Stochastic processes and their applications},
  volume={119},
  number={7},
  pages={2249--2276},
  year={2009},
  publisher={Elsevier}
}

@article{jing2014estimation,
  title={On the estimation of integrated volatility with jumps and microstructure noise},
  author={Jing, Bing-Yi and Liu, Zhi and Kong, Xin-Bing},
  journal={Journal of Business \& Economic Statistics},
  volume={32},
  number={3},
  pages={457--467},
  year={2014},
  publisher={Taylor \& Francis}
}

@article{kristensen2010nonparametric,
  title={Nonparametric filtering of the realized spot volatility: A kernel-based approach},
  author={Kristensen, Dennis},
  journal={Econometric Theory},
  volume={26},
  number={1},
  pages={60--93},
  year={2010},
  publisher={Cambridge University Press}
}

@article{large2011estimating,
  title={Estimating quadratic variation when quoted prices change by a constant increment},
  author={Large, Jeremy},
  journal={Journal of Econometrics},
  volume={160},
  number={1},
  pages={2--11},
  year={2011},
  publisher={Elsevier}
}

@article{li2018unified,
 title = {A unified approach to volatility estimation in the presence of both rounding and random market microstructure noise},
 author={Li, Yingying and Zhang, Zhiyuan and Li, Yichu},
 journal = {Journal of Econometrics},
 volume = {203},
 number = {2},
 pages = {187-222},
 year = {2018},
}

@article{li2021volatility,
  title={Volatility of volatility: Estimation and tests based on noisy high frequency data with jumps},
  author={Li, Yingying and Liu, Guangying and Zhang, Zhiyuan},
  journal={Journal of Econometrics},
  year={2021},
  publisher={Elsevier}
}

@article{li2022reading,
  title={Reading the Candlesticks: An OK Estimator for Volatility},
  author={Li, Jia and Wang, Dishen and Zhang, Qiushi},
  journal={The Review of Economics and Statistics},
  pages={1--45},
  year={2022}
}

@article{malliavin2002fourier,
  title={Fourier series method for measurement of multivariate volatilities},
  author={Malliavin, Paul and Mancino, Maria Elvira},
  journal={Finance and Stochastics},
  volume={6},
  number={1},
  pages={49--61},
  year={2002},
  publisher={Springer}
}

@article{malliavin2009fourier,
  title={A Fourier transform method for nonparametric estimation of multivariate volatility},
  author={Malliavin, Paul and Mancino, Maria Elvira},
  journal={The Annals of Statistics},
  volume={37},
  number={4},
  pages={1983--2010},
  year={2009},
  publisher={Institute of Mathematical Statistics}
}

@article{mancini2015spot,
  title={Spot volatility estimation using delta sequences},
  author={Mancini, Cecilia and Mattiussi, Vanessa and Ren{\`o}, Roberto},
  journal={Finance and Stochastics},
  volume={19},
  number={2},
  pages={261--293},
  year={2015},
  publisher={Springer}
}

@article{mancino2015fourier,
  title={Fourier spot volatility estimator: Asymptotic normality and efficiency with liquid and illiquid high-frequency data},
  author={Mancino, Maria Elvira and Recchioni, Maria Cristina},
  journal={PloS one},
  volume={10},
  number={9},
  pages={e0139041},
  year={2015},
  publisher={Public Library of Science San Francisco, CA USA}
}

@article{mancino2008robustness,
  title={Robustness of Fourier estimator of integrated volatility in the presence of microstructure noise},
  author={Mancino, Maria Elvira and Sanfelici, Simona},
  journal={Computational Statistics \& data analysis},
  volume={52},
  number={6},
  pages={2966--2989},
  year={2008},
  publisher={Elsevier}
}

@article{patton2011volatility,
  title={Volatility forecast comparison using imperfect volatility proxies},
  author={Patton, Andrew J},
  journal={Journal of Econometrics},
  volume={160},
  number={1},
  pages={246--256},
  year={2011},
  publisher={Elsevier}
}

@article{ogawa2008real,
  title={Real-time scheme for the volatility estimation in the presence of microstructure noise},
  author={Ogawa, Shigeyoshi},
  journal={Monte Carlo Methods and Applications},
  year=2008,
  volume={14},
  number={4},
  pages={331-342},
}

@article{ogawa2011improved,
  title={An improved two-step regularization scheme for spot volatility estimation},
  author={Ogawa, Shigeyoshi and Sanfelici, Simona},
  journal={Economic Notes},
  volume={40},
  number={3},
  pages={107--134},
  year={2011},
  publisher={Wiley Online Library}
}

@article{robert2011new,
  title={A new approach for the dynamics of ultra-high-frequency data: The model with uncertainty zones},
  author={Robert, Christian Y and Rosenbaum, Mathieu},
  journal={Journal of Financial Econometrics},
  volume={9},
  number={2},
  pages={344--366},
  year={2011},
  publisher={Oxford University Press}
}

@article{roll1984simple,
  title={A simple implicit measure of the effective bid-ask spread in an efficient market},
  author={Roll, Richard},
  journal={The Journal of finance},
  volume={39},
  number={4},
  pages={1127--1139},
  year={1984},
  publisher={Wiley Online Library}
}

@article{smith2003statistical,
  title={Statistical theory of the continuous double auction},
  author={Smith, Eric and Farmer, J Doyne and Gillemot, L{\'a}szl{\'o} and Krishnamurthy, Supriya},
  journal={Quantitative finance},
  volume={3},
  number={6},
  pages={481},
  year={2003},
  publisher={IOP Publishing}
}

@article{vortelinos2014optimally,
  title={Optimally sampled realized range-based volatility estimators},
  author={Vortelinos, Dimitrios I},
  journal={Research in International Business and Finance},
  volume={30},
  pages={34--50},
  year={2014},
  publisher={Elsevier}
}

@article{zhang2006efficient,
 author = {Zhang, Lan},
 title = {{Efficient estimation of stochastic volatility using noisy observations: a multi-scale approach}},
 volume = {12},
 journal = {Bernoulli},
 number = {6},
 publisher = {Bernoulli Society for Mathematical Statistics and Probability},
 pages = {1019 -- 1043},
 year = {2006},
}

@article{zhang2005tale,
  title={A tale of two time scales: Determining integrated volatility with noisy high-frequency data},
  author={Zhang, Lan and Mykland, Per A and A{\"i}t-Sahalia, Yacine},
  journal={Journal of the American Statistical Association},
  volume={100},
  number={472},
  pages={1394--1411},
  year={2005},
  publisher={Taylor \& Francis}
}

@article{zhou1996high,
  title={High-frequency data and volatility in foreign-exchange rates},
  author={Zhou, Bin},
  journal={Journal of Business \& Economic Statistics},
  volume={14},
  number={1},
  pages={45--52},
  year={1996},
  publisher={Taylor \& Francis}
}

@article{zu2014estimating,
  title={Estimating spot volatility with high-frequency financial data},
  author={Zu, Yang and Boswijk, H Peter},
  journal={Journal of Econometrics},
  volume={181},
  number={2},
  pages={117--135},
  year={2014},
  publisher={Elsevier}
}

@article{lilloFarmer2004,
  title={The long memory of the efficient market},
  author={Lillo, Fabrizio and Farmer, J Doynne},
  journal={Studies in nonlinear dynamics \& econometrics},
  year={2004},
  volume={8},
  pages={3}
}

@article{eisler,
  title={The price impact of order book events: Market orders, limit orders and cancellations},
  author={Eisler, Zolt\'an. and Bouchaud, Jean-Philippe and Kockelkoren, Julien},
  journal={Quantitative Finance},
  volume={12},
   pages={1395-1419},
  year={2012}
}

\newpage
\appendix
\section*{Appendix A}\label{appendixmat}

\subsection*{A.1 Additional performance results (variable $\theta$ and $\theta^{reinit}$)}

\subsubsection*{A.1.1 5 regimes}
\begin{table}[htbp!]
\begin{center}
    \begin{tabular}{c|c|c|c|c|c|c|c}
        \hline\hline
        \multicolumn{8}{c}{Integrated variance estimators - relative bias}\\
		\hline\hline
		Estimator & mid-price & rank & micro-price & rank & trade-price & rank &
		av. rank\\
		\hline
		Pre-averaging & 0.02365 & 1 & 0.02519 & 1 &  0.02364 & 1 & 1\\
		Med RV & -0.03787  & 2 & -0.03778 & 2 & -0.03738 & 2 & 2\\
		Min RV & -0.03828 & 3 &-0.03821 & 3 & -0.03782 & 3 & 3\\
		Range &  -0.07194 & 4 & -0.05015 & 4 & -0.06590 & 4 & 4\\
		Unified  & 0.092375 & 5 & 0.08478 & 5 & 0.16468 & 7&6\\
		Fourier & 0.12680 & 7 & 0.10451 & 6 & 0.12580 & 6 & 6.33 \\
		Alternation & 0.09803  & 6 & 0.19524  & 12 & 0.09607 & 5 & 7.66\\
		Kernel & 0.17164  & 9 & 0.14899 & 8 & 0.31278 & 8 & 8.33\\
		Maximum likelihood & 0.16986   & 8  & 0.14813 & 7 & 0.41152 & 11 & 8.66\\
		Two-scale RV & 0.17577  & 10 & 0.15552 & 9 & 0.41140 &  9 & 9.33\\
		Multi-scale RV & 0.17592 & 11 & 0.15569 &10 & 0.41152 &10 & 10.33\\
		Bias-corrected RV & 0.17606 & 12 & 0.15583  & 11& 0.41164 & 12 & 11.66 
    \end{tabular}
    \caption{Performance and ranking of the integrated variance estimators for the series of mid-price, micro-price and trade-price with 5 regimes of $\theta$ and $\theta^{\text{reinit}}$, according to the relative bias. The average ranking is reported in the last column.}
	\label{tab:integratedbias5l}
\end{center}
\end{table}

\begin{table}[htbp!]
\begin{center}
    \begin{tabular}{c|c|c|c|c|c|c|c}
        \hline\hline
        \multicolumn{8}{c}{Integrated variance estimators - relative MSE}\\
		\hline\hline
		Estimator & mid-price & rank & micro-price & rank & trade-price & rank &
		av. rank\\
		\hline
		Unified  & 0.01002 & 1  & 0.00864 & 1 & 0.02881 & 7 & 3\\
		Fourier & 0.01525 & 3 & 0.01251 & 2 & 0.01823 & 4 & 3\\
		Range &  0.01769 & 4  & 0.01253 & 3 & 0.01029 & 2 & 3\\
		Pre-averaging & 0.01888 & 5 & 0.01691 & 4 & 0.01685  & 3 & 4\\
		Alternation & 0.01020  & 2 &  0.03927 & 12 & 0.00332 & 1 & 5\\
		Med RV & 0.02022 & 6 & 0.02022 & 5 &  0.02010 & 5 & 5.33\\
		Min RV & 0.02511 & 7 & 0.02510 & 8 & 0.02492 & 6 & 7\\
		Maximum likelihood & 0.03008  & 8 & 0.02311 & 6 & 0.12163 & 9 & 7.66\\
		Kernel & 0.03075  & 9 & 0.02334  & 7 & 0.10014 & 8 & 8\\
		Two-scale RV & 0.03211 & 10 & 0.02534 & 9 & 0.17089 & 10 & 9.66\\
		Multi-scale RV & 0.032161 & 11 & 0.02539 & 10 & 0.17098 &11 & 10.66\\
		Bias-corrected RV & 0.03222 & 12 & 0.02543  & 11 & 0.17102 & 12& 11.66\\
    \end{tabular}
    \caption{Performance and ranking of the integrated variance estimators for the series of mid-price, micro-price and trade-price with 5 regimes of $\theta$ and $\theta^{\text{reinit}}$, according to the relative mean square error. The average ranking is reported in the last column.}
	\label{tab:integratedmse5l}
\end{center}
\end{table}

\begin{table}[htbp!]
\begin{center}
    \begin{tabular}{c|c|c|c|c|c|c|c}
        \hline\hline
        \multicolumn{8}{c}{Spot variance estimators - relative integrated bias}\\
		\hline\hline
		Estimator & mid-price & rank & micro-price & rank & trade-price & rank &
		av. rank\\
		\hline
        Fourier & -0.00099  & 1 & -0.00083 &1 & -0.00597 & 1&1 \\
        Regularized & -0.00904 & 2& -0.00912 &2 & -0.00877 &2 & 2 \\
        Pre-averaging kernel & -0.06324 &3 & -0.04842 & 3& -0.05288 & 3& 3 \\
        Optimal candlestick & -0.16978 &4 & -0.12938 &4 & -0.07699 &4 & 4\\
        Pre-averaging & 0.17632 & 5& 0.17629 &5 & -0.11810 & 5&5 \\
        Kernel & 0.20633 &6 & 0.26122 &6 & 0.27361 & 6&6 \\
        Two-scale & -0.24749  &7 & 0.26543 & 7& 0.81305 &7 &7 \\
    \end{tabular}
    \caption{Performance and ranking of the spot variance estimators for the series of mid-price, micro-price and trade-price with 5 regimes of $\theta$ and $\theta^{\text{reinit}}$, according to the relative integrated bias. The average ranking is reported in the last column.}
	\label{tab:spotbias5l}
\end{center}
\end{table}

\begin{table}[htbp!]
\begin{center}
    \begin{tabular}{c|c|c|c|c|c|c|c}
        \hline\hline
        \multicolumn{8}{c}{Spot variance estimators - relative integrated MSE}\\
		\hline\hline
		Estimator & mid-price & rank & micro-price & rank & trade-price & rank &
		av. rank\\
		\hline
        Fourier & 0.035801  &1 & 0.03585 & 1& 0.03616 &1 & 1\\
        Pre-averaging kernel & 0.19077 &2 & 0.19206 &2 & 0.14159 &2 &2 \\
        Regularized & 0.21590 & 3& 0.21592 &3 & 0.21582 &3 &3 \\
        Optimal candlestick &0.31354 &4 & 0.31110&4 &0.32879 &4 &4 \\
        Kernel & 0.42691 &5 & 0.49152 &5 & 0.82784 & 5& 5\\
        Two-scale & 1.62405  &6 & 1.61549 &6 & 2.38642 & 6&6 \\
        Pre-averaging & 2.52237  &7 & 2.52255 & 7& 2.52422 &7 &7 \\
    \end{tabular}
    \caption{Performance and ranking of the spot variance estimators for the series of mid-price, micro-price and trade-price with 5 regimes of $\theta$ and $\theta^{\text{reinit}}$, according to the relative integrated mean square error. The average ranking is reported in the last column.}
	\label{tab:spotmse5l}
\end{center}
\end{table}

\newpage
\subsubsection*{A.1.2 10 regimes}

\begin{table}[htbp!]
\begin{center}
    \begin{tabular}{c|c|c|c|c|c|c|c}
        \hline\hline
        \multicolumn{8}{c}{Integrated variance estimators - relative bias}\\
		\hline\hline
		Estimator & mid-price & rank & micro-price & rank & trade-price & rank &
		av. rank\\
		\hline
		Med RV & 0.06614 & 1 & 0.06609 & 1 &  0.06660 & 1 & 1\\
		Pre-averaging & 0.07184 & 2 & 0.07182 & 2 & 0.07237 & 2 & 2\\
		Min RV & 0.07669 & 3 & 0.07667 & 3 & 0.07680  & 4 & 3.33\\
		Range &  -0.12325 & 5  & -0.11637 & 5 & 0.13146 & 5 & 5\\
		Unified  & 0.12626 & 6  & 0.11419 & 4 &  0.19885 & 7 & 5.66\\
		Fourier & 0.16911 & 7 & 0.14236 & 6 & 0.17286 & 6 & 6.33\\
		Alternation & 0.09319  & 4 &  0.16445 & 12 & -0.0741 & 3 & 6.33\\
		Maximum likelihood & 0.18492  & 8 & 0.15720 & 7 & 0.37057 & 9 & 8\\
		Kernel  & 0.18758  & 10 & 0.15928  & 8 & 0.34518 & 8 & 9\\
		Two-scale RV & 0.18740 & 9 &  0.16030 & 9 &  0.43099 & 10 & 9.33\\
		Multi-scale RV & 0.18759 & 11 & 0.16048 & 10 &  0.43118 &11 & 10.66\\
		Bias-corrected RV &  0.18773 & 12 & 0.16061  & 11 & 0.43131 & 12& 11.66\\
    \end{tabular}
    \caption{Performance and ranking of the integrated variance estimators for the series of mid-price, micro-price and trade-price with 10 regimes of $\theta$ and $\theta^{\text{reinit}}$, according to the relative bias. The average ranking is reported in the last column.}
	\label{tab:integratedbias10l}
\end{center}
\end{table}

\begin{table}[htbp!]
\begin{center}
    \begin{tabular}{c|c|c|c|c|c|c|c}
        \hline\hline
        \multicolumn{8}{c}{Integrated variance estimators - relative MSE}\\
		\hline\hline
		Estimator & mid-price & rank & micro-price & rank & trade-price & rank &
		av. rank\\
		\hline
		Unified  & 0.01742 & 2  & 0.01447 & 1 & 0.04113 & 7 & 3.33\\
		Fourier & 0.02189 & 3 & 0.02270 & 2 & 0.03208 & 5 & 3.33\\
		Range &  0.02660 & 4  & 0.02659 & 4 & 0.02070 & 2 & 3.33\\
		Alternation & 0.01003  & 1 &  0.02796 & 11 & 0.00111 & 1 & 4.33\\
		Pre-averaging & 0.02732 & 5 & 0.02632 & 5 & 0.02734  & 3 & 4.33\\
		Maximum likelihood & 0.03541  & 7 & 0.02586 & 3 & 0.13900 & 9 & 6.33\\
		Med RV & 0.02881 & 6 & 0.02712 & 10 &  0.02987 & 4 & 6.66\\
		Kernel  & 0.03641  & 10 & 0.02655  & 6 & 0.12112 & 8 & 8\\
		Min RV & 0.03572 & 8 & 0.03674 & 12 & 0.03674 & 6 & 8.66\\
		Multi-scale RV & 0.03639 & 9 & 0.02689 & 8 & 0.18767 &10 & 9\\
		Two-scale RV & 0.03642 & 11 & 0.02684 & 7 & 0.18770 & 11 & 9.66\\
		Bias-corrected RV & 0.03645 & 12 & 0.02694  & 9 & 0.18779 & 12& 11\\
    \end{tabular}
    \caption{Performance and ranking of the integrated variance estimators for the series of mid-price, micro-price and trade-price with 10 regimes of $\theta$ and $\theta^{\text{reinit}}$, according to the relative mean square error. The average ranking is reported in the last column.}
	\label{tab:integratedmse10l}
\end{center}
\end{table}

\begin{table}[htbp!]
\begin{center}
    \begin{tabular}{c|c|c|c|c|c|c|c}
        \hline\hline
        \multicolumn{8}{c}{Spot variance estimators - relative integrated bias}\\
		\hline\hline
		Estimator & mid-price & rank & micro-price & rank & trade-price & rank &
		av. rank\\
		\hline
        Fourier &  -0.02342  &1 & -0.02412 & 1& -0.01745 &1 & 1\\
        Regularized & -0.02472 & 2& -0.02471 &2 & -0.02438 &2 &2 \\
        Pre-averaging kernel & -0.08816 &3 & -0.09235 &3 & -0.07796 &3 & 3 \\
        Optimal candlestick & -0.20641 &5 & -0.15948 &4 & -0.0838 &4 &4.33 \\
        Pre-averaging & 0.24979  &6 &  0.24976 & 5 &  0.24958 &5 & 5.33 \\
        Kernel & 0.20568 &4 & 0.25983 &6 & 0.79978 & 6& 5.66\\
        Two-scale & -0.27408  &7 & -0.28400 &7 & -0.28864 & 7& 7 \\
        
    \end{tabular}
    \caption{Performance and ranking of the spot variance estimators for the series of mid-price, micro-price and trade-price with 10 regimes of $\theta$ and $\theta^{\text{reinit}}$, according to the relative integrated bias. The average ranking is reported in the last column.}
	\label{tab:spotbias10l}
\end{center}
\end{table}

\begin{table}[htbp!]
\begin{center}
    \begin{tabular}{c|c|c|c|c|c|c|c}
        \hline\hline
        \multicolumn{8}{c}{Spot variance estimators - relative integrated MSE}\\
		\hline\hline
		Estimator & mid-price & rank & micro-price & rank & trade-price & rank &
		av. rank\\
		\hline
        Fourier &  0.05523  &1 & 0.05527 & 1& 0.05618 &1 & 1\\
        Pre-averaging kernel & 0.19695 &2 & 0.19833 &2 & 0.15153 &2 &2 \\
        Regularized & 0.23490 & 3& 0.23502 &3 & 0.23458 &3 &3 \\
        Optimal candlestick & 0.31762 &4 & 0.31529 &4 & 0.34982 &4 &4\\
        Kernel & 0.57819 &5 & 0.68831 &5 & 0.79963 & 5& 5\\
        Two-scale & 1.61446  &6 & 1.65113 &6 & 2.78745 & 7& 6.33 \\
        Pre-averaging & 2.49265  &7 &  2.49293 & 7& 2.49131 &6 & 6.66
    \end{tabular}
    \caption{Performance and ranking of the spot variance estimators for the series of mid-price, micro-price and trade-price with 10 regimes of $\theta$ and $\theta^{\text{reinit}}$, according to the relative integrated mean square error. The average ranking is reported in the last column.}
	\label{tab:spotmse10l}
\end{center}
\end{table}

\subsection*{A.2 Sample daily volatility trajectories of spot variance estimates (micro-price and trade-price)}

\begin{figure}[htbp!]
    \centering
    \includegraphics[width =1\linewidth]{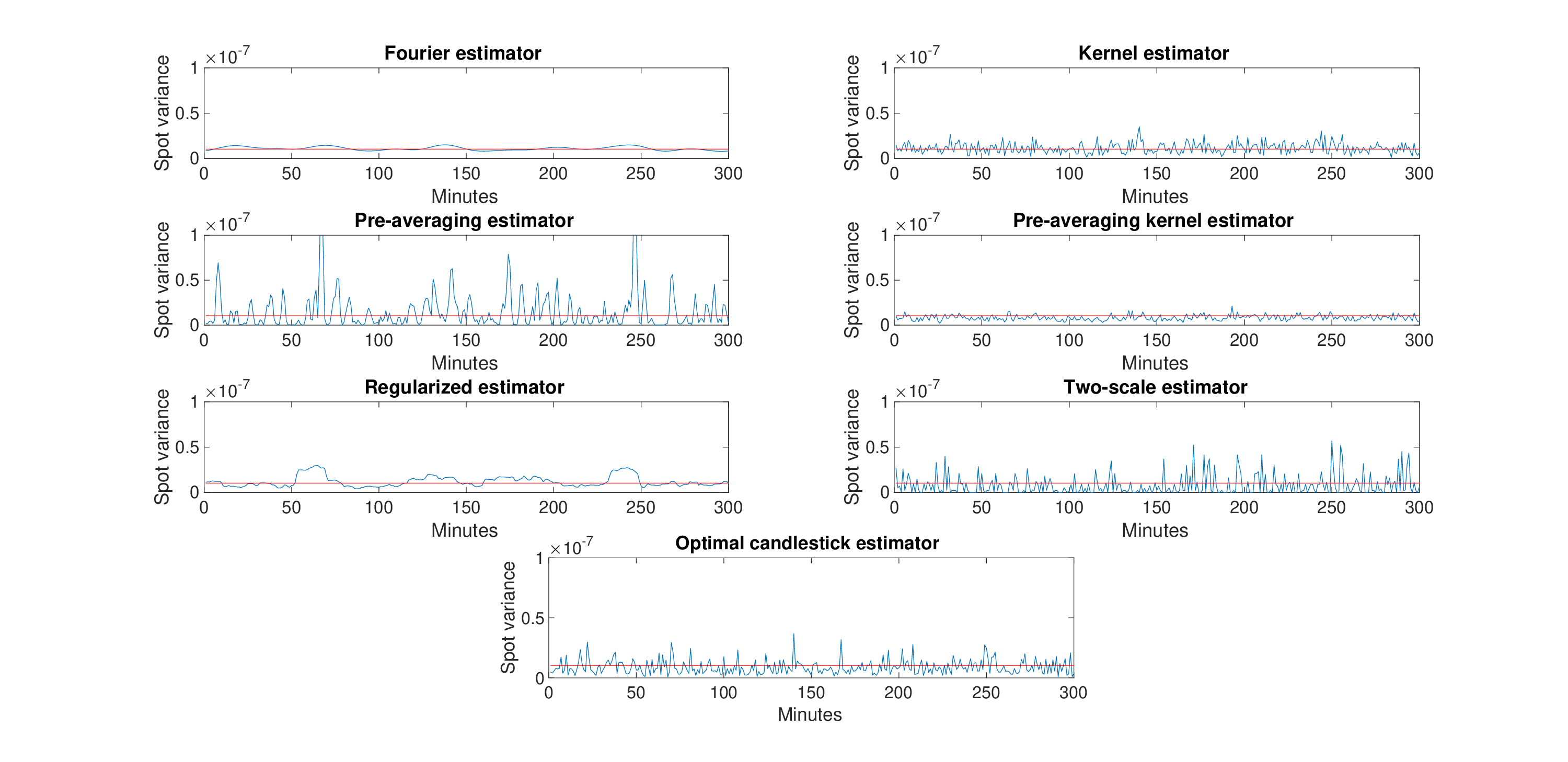}
    \caption{Constant $\theta$ and $\theta^{\text{reinit}}$: sample trajectories of spot variance estimators computed from micro-prices (in blue) and true volatility (in red).}
    \label{fig:spotmicro}
\end{figure}

\begin{figure}[htbp!]
    \centering
    \includegraphics[width =1\linewidth]{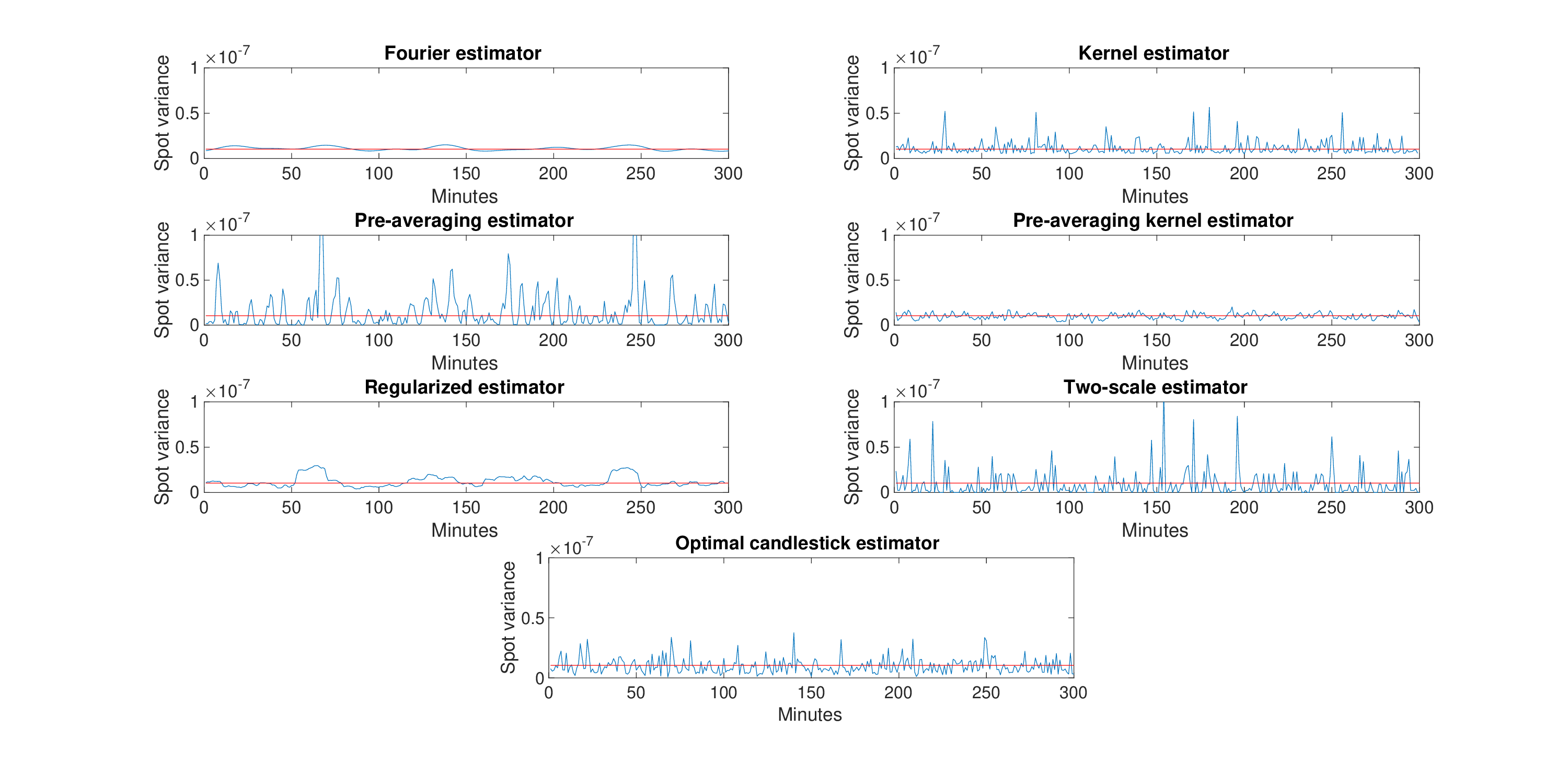}
    \caption{Constant $\theta$ and $\theta^{\text{reinit}}$: sample trajectories of spot variance estimators computed from trade-prices (in blue) and true volatility (in red).}
    \label{fig:spottrade}
\end{figure}

\begin{figure}[htbp!]
    \centering
    \includegraphics[width =1\linewidth]{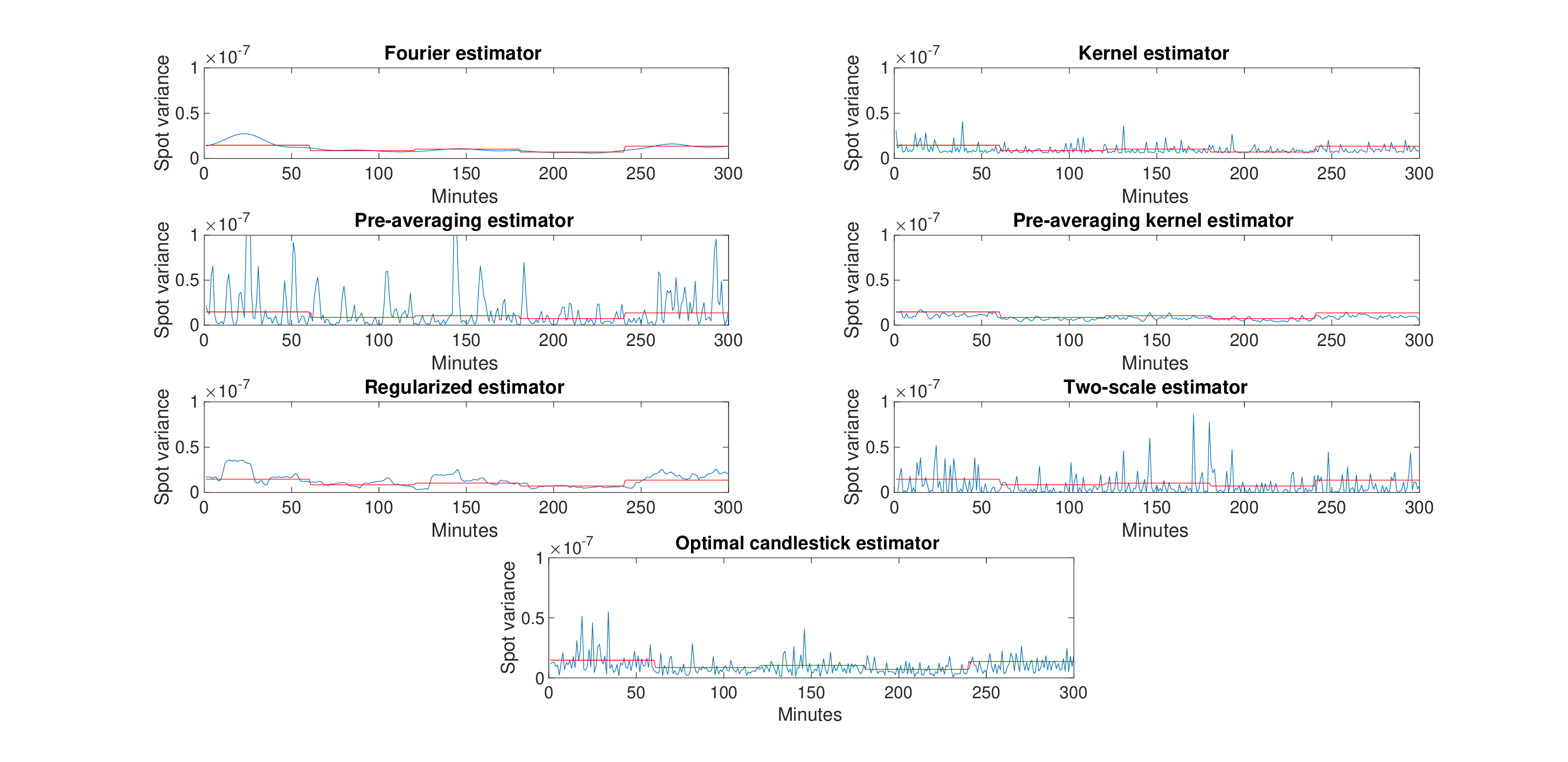}
    \caption{Variable $\theta$ and $\theta^{\text{reinit}}$ (5 regimes): sample trajectories of spot variance estimators computed from micro-prices (in blue) and true volatility (in red).}
    \label{fig:spotmicro5lv}
\end{figure}

\begin{figure}[htbp!]
    \centering
    \includegraphics[width =1\linewidth]{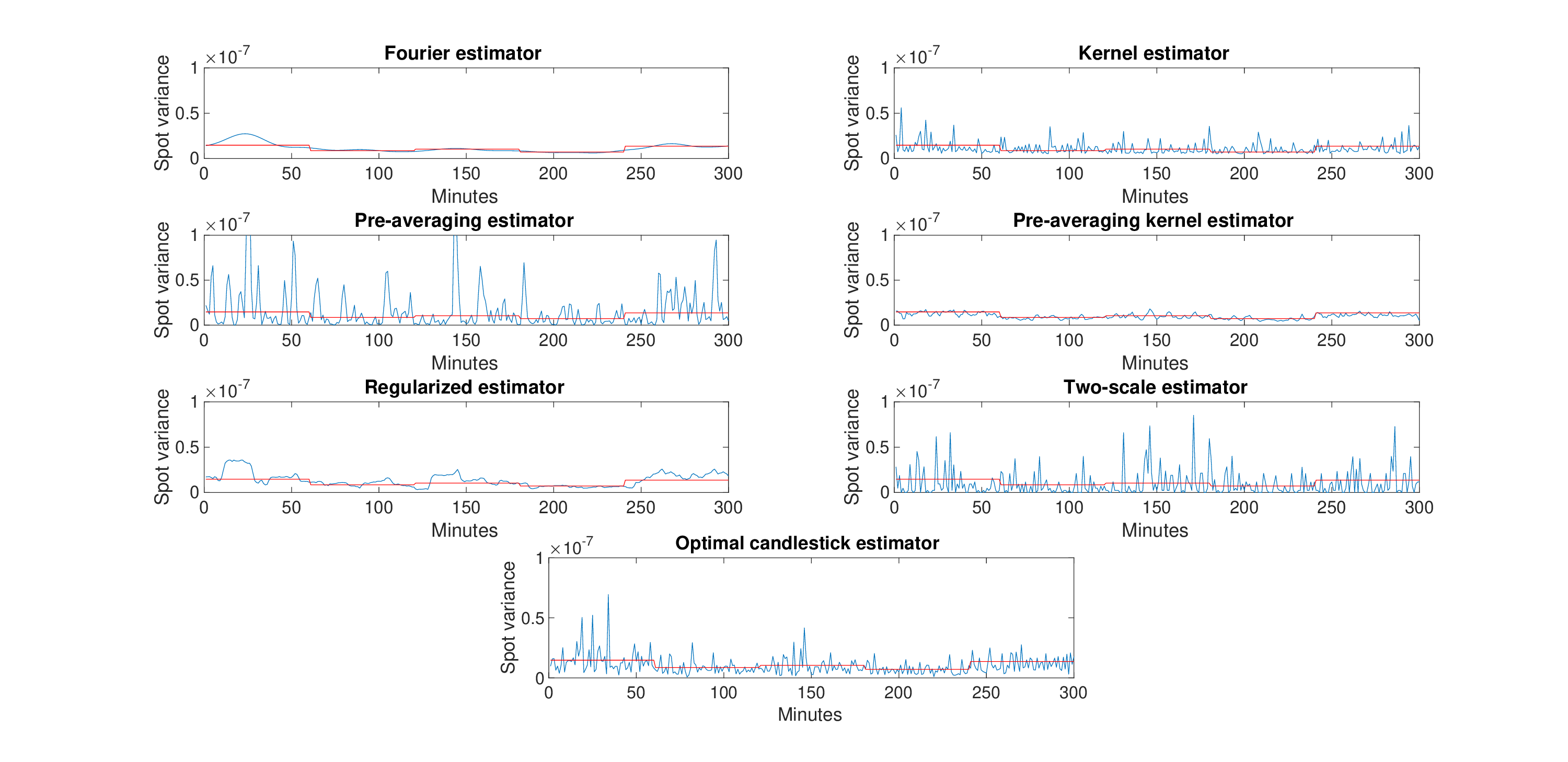}
    \caption{Variable $\theta$ and $\theta^{\text{reinit}}$ (5 regimes): sample trajectories of spot variance estimators computed from trade-prices (in blue) and true volatility (in red).}
    \label{fig:spottrade5lv}
\end{figure}

\begin{figure}[htbp!]
    \centering
     \includegraphics[width =1\linewidth]{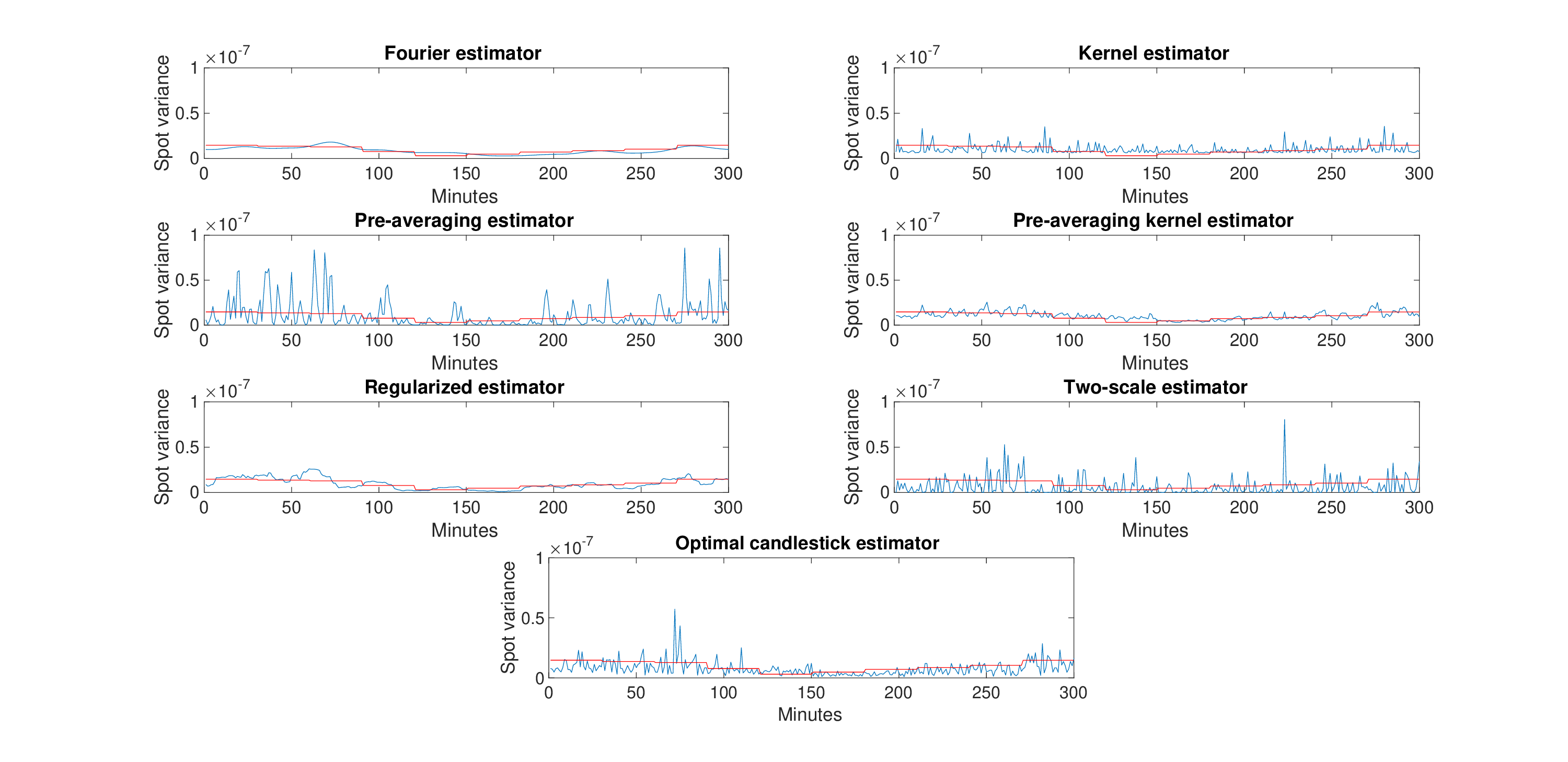}
    \caption{Variable $\theta$ and $\theta^{\text{reinit}}$ (10 regimes): sample trajectories of spot variance estimators computed from micro-prices (in blue) and true volatility (in red).}
    \label{fig:spotmicro10lv}
\end{figure}

\begin{figure}[htbp!]
    \centering
     \includegraphics[width =1\linewidth]{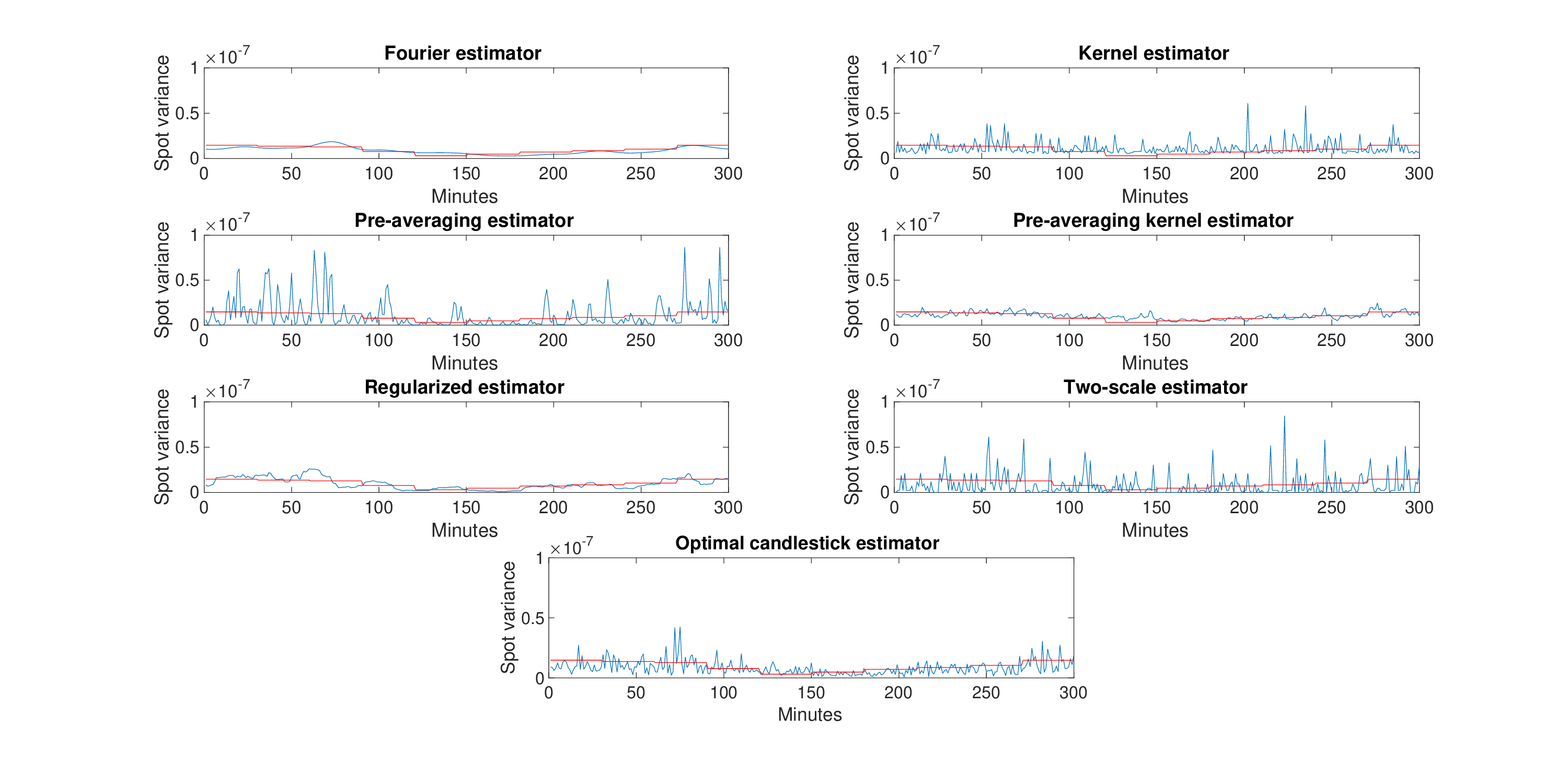}
    \caption{Variable $\theta$ and $\theta^{\text{reinit}}$ (10 regimes): sample trajectories of spot variance estimators computed from trade-prices (in blue) and true volatility (in red).}
    \label{fig:spottrade10lv}
\end{figure}

\end{document}